\documentclass[aps,twocolumn]{revtex4-1}
\usepackage{amsmath,amssymb}
\usepackage{slashed}
\usepackage{graphicx}
\usepackage{bm}
\usepackage[T1]{fontenc}
\usepackage[utf8]{inputenc}
\usepackage{gauss} 
\usepackage[top=1.0in,bottom=1.0in,left=1.0in,right=1.0in]{geometry}
\usepackage[colorlinks,linkcolor=blue,citecolor=blue]{hyperref}
\newcommand{\be}{\begin{equation}}
\newcommand{\ee}{\end{equation}}
\newcommand{\bea}{\begin{eqnarray}}
\newcommand{\eea}{\end{eqnarray}}
\newcommand{\ba}{\begin{eqnarray}}
\newcommand{\ea}{\end{eqnarray}}

\begin{document}

\title{
Bridging hadronic and vacuum structure
by heavy quarkonia
}

\author{Nicholas Miesch}
\email{nicholas.miesch@stonybrook.edu}
\affiliation{Center for Nuclear Theory, Department of Physics and Astronomy, Stony Brook University, Stony Brook, New York 11794--3800, USA}

\author{Edward Shuryak}
\email{edward.shuryak@stonybrook.edu}
\affiliation{Center for Nuclear Theory, Department of Physics and Astronomy, Stony Brook University, Stony Brook, New York 11794--3800, USA}

\author{Ismail Zahed}
\email{ismail.zahed@stonybrook.edu}
\affiliation{Center for Nuclear Theory, Department of Physics and Astronomy, Stony Brook University, Stony Brook, New York 11794--3800, USA}

\begin{abstract} 
We discuss the central and, mostly, spin-dependent potentials in heavy quarkonia $\bar b b, \bar c c$, with two goals in mind. The first is phenomenological: using the splitting between the 1S and 2S pairs, as well as the 1P and 2P quartet masses,  we obtain
very  accurate values of all matrix elements of the spin-depepdent potentials. The second is theoretical: using standard wave functions,  we compute these matrix elements, from perturbative, nonperturbative and ``string" contributions. The model for nonperturbative effects is  a  ``dense instanton liquid model", in which the QCD vacuum is made of (strongly color correlated) instanton-antiinstanton pairs or "molecules". We calcuate their effect via
standard Wilson lines, with or without extra powers of gauge fields. We find that this model  provides a reasonable description of all central, spin-spin and spin-orbit forces at
distances $r= 0-0.7 \, fm $, relevant for $\bar b b $ and $\bar c c$ quarkonia.
 \end{abstract}

\maketitle

\section{Introduction}
\subsection{Topological solitons in the  QCD vacuum  }

The mainstream of studies of the vacuum structure in QCD (and related gauge theories) is today centered on numerical simulations, using
lattice formulation in 4d Euclidean space-time. While its history
span nearly 50 years, since  Wilson's formulation, only recently a number of  technical
goals have been reached, such as quantitative simulations with quarks light enough to make chiral extrapolations unnecessary.
This led to spectacular results in reproducing many properties
of hadrons -- masses and multiple matrix elements, too many to
mention here.
Yet it remains imperative that a ``black box" approach based on  numerical simulations, be complemented by better understanding of the underlying dynamics. 

 Nonabelian gauge fields -- the main component of the Standard Model - possess  configurations with a variety
of topological properties. They include 2d ``vortices", 3d ``monopoles" and 4d "instantons". They are all inter-related, e.g. monopoles correspond to crossing of vortices, etc. 
Their presence in the vacuum gauge fields  have been found and quantified using
lattice  configurations. Different methods -- such as e.g.
the renormalization group (RG) based on gradient flow -- allows the localization of the topological objects, and their separation  from the perturbative gluons,
in a well-controlled way. 
Their densities and interactions were shown to be directly related to key
parameters of hadronic spectroscopy. (E.g. the density of center vortices
define the string tension of the QCD flux tubes.)

Remarkable pictures of the ``vacuum fields snapshots" from some of these numerical simulations~\cite{Ilgenfritz:2008ia,Biddle:2019gke,Mickley:2023exg},
 have revealed topological structure of the vacuum fields, including 2d  vortices, 3d monopoles  and {\em 4d instantons} ($I$). 
There is a very significant practical difference between these structures.  While vortices and monopoles are defined by topology alone,  instantons are true solitons, classical solutions  of (Euclidean) Yang-Mills equations \cite{Belavin:1975fg}. 
We restarted the
attempts to describe them in terms of interacting ensembles of  topological solitons. This trend is similar to e.g. describing turbulence in various systems via 
ensembles of  individual ``nonlinear waves", or solitons.  

A while ago, using phenomenological and theoretical arguments,
one of us \cite{Shuryak:1981fza} argued that in the QCD vacuum, the 
instanton size distribution is strongly peaked at a certain value,  so that their action is large
\be 
\rho \approx {1 \over 3} \, {\rm fm},\qquad {S_{inst} \over \hbar }\sim O(10)  \gg 1
\ee
If so, it should be possible to develop a {\em semiclassical theory} of
an instanton ensemble. Numerical studies of such ensemble
have been done in the 1980s and 1990s, for a summary see \cite{Schafer:1996wv}. 

Already the simplest ''dilute instanton liquid model" (ILM),
with uncorrelated instantons at density 
\be \label{eqn_n_ILM}
n_{ILM} \approx 1 \, {\rm fm}^{-4} \ee 
as predicted in~\cite{Shuryak:1981fza}, was found to
describe  a number of phenomena, including the 
explicit breakings of $U(1)_A$ (related to the $\eta'$) and  the spontaneous $SU(N_f)$ chiral symmetry breaking (and the properties of $\pi,K,\eta $). A key to all these phenomena is the four- (or six-) fermion 't Hooft effective Lagrangian \cite{tHooft:1976snw}, generated by the fermionic zero modes.

At the same time, statistical studies of interacting instantons 
revealed strong instanton-antiinstanton ($I\bar I$) correlations.
A quantitative theory of such "molecules"--incomplete tunneling events-- have not been carried out then, although later they were related to sphaleron production, in electroweak theory and QCD. 

The $I\bar I$ correlated pairs were seen on the lattice since 1990's. With various ``cooling" methods, as well as more modern 
``gradient flow" \cite{Athenodorou:2018jwu}, one can see
them after very little cooling. With increasing cooling time, one further observes how they get annihilated,  converging eventually to 
a dilute set of well-separated instantons, with the ILM parameters. The most important
observation is that the density of ``molecules" is about an order
of magnitude larger than that ofthe  dilute ILM (\ref{eqn_n_ILM}).

The dynamical contribution of the correlated  $I\bar I$ pairs also resurfaced 
few years ago in several   applications, where the
nonperturbative gauge fields (rather than the fermionic ones)  are directly involved. The first 
was our study of the pion form factors in the semi-hard regime\cite{Shuryak:2020ktq} .  The second  was our study of the heavy quark static and spin-dependent potentials~\cite{Shuryak:2021fsu}. In both analyses,
 the contribution of   these $I \bar I$ ``molecules" was in line with phenomenological expectations, provided
  the molecular density is as large as seen on the lattice (in the zero cooling time limit). We called the corresponding model a ``dense instanton liquid model" (DILM).  
  
In this paper we use the same model and
continue ``bridging the gap" between vacuum and hadronic structures, by focusing on the relation between
 nonperturbative vacuum fields and quarkonium potentials. In such testing ground    light quarks, chiral symmetry breaking, pions or 't Hooft Lagrangian cannot play any role. The relation itself is well known: the central potential
is related to  the Wilson loop, while the spin-dependent potentials follow by dressing the loop  by magnetic and electric fields.

\begin{figure}
    \centering
    \includegraphics[width=6cm]{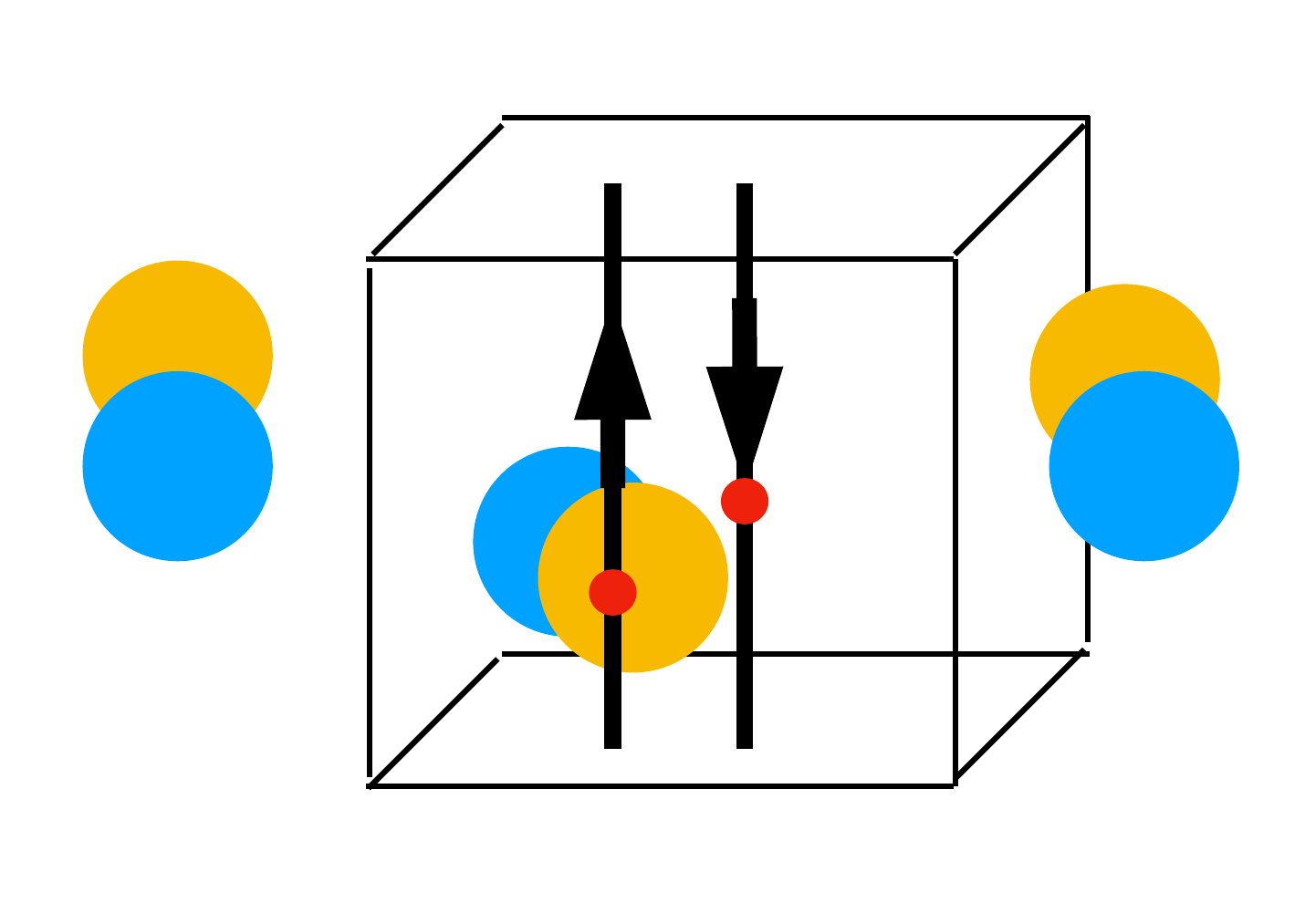}
    \caption{Schematic snapshot of the QCD vacuum fields, with the insertion of two Wilson lines. The blue and yellow blobs are regions of positive and negative
    topological charges (instanton and anti-instanton-like), grouped into strongly correlated ``molecules". The cube indicates a mean volume per ``molecule". The small red dots show the locations of the magnetic and electric field strengths insertions on the Wilson lines, to measure
    the spin-dependent forces.  }
    \label{fig_cube}
\end{figure}

The masses and  wave functions of S- and P-shell
states, are under good theoretical control, 
which allows for the evaluation of the matrix elements
of the spin-spin, spin-orbit and tensor forces from data. These empirical values are then compared to the calculated perturbative and -- most importantly -- the nonperturbative contributions. 
All we do is evaluating
 ``Wilson lines", with and without additional field strengths,
using fields of instantons or their correlated pairs or ``molecules".  This leads to nontrivial correlations of  the gauge potentials $A_\mu$,
with electric and magnetic fields at certain locations. The setting is schematically shown in fig.\ref{fig_cube}. It is basically the same as done (or can be done) using lattice gauge configurations.

This paper is structured as follows. The remainder of the introduction includes the subsections \ref{sec_tunneling},
where we briefly recall the concept of  $tunneling$ via  topological barriers (instantons), as well as that of ``incomplete tunneling" or $\bar I I $ ``molecules".
The general relations between the vacuum fields and the static potentials are recalled in section~\ref{sec_potentialsX}.  In section \ref{sec_ratio} we provide more details on the gauge
configuration used for molecules, and in section~\ref{sec_potentials} we analyze their contribution to the central potential.
The main  section in  this work is \ref{sec_spin-spin},
where the focus is on the spin-dependent potentials. 
We use the existing empirical information on S- and P-shell for quarkonia, to extract the values of the matrix elements for the spin-spin forces (S-shell), and spin-spin, spin-orbit and tensor forces (P-shell). 
We also evaluate the same matrix elements, using the
perturbative one-gluon, the instanton and the molecular
contributions. In  section \ref{sec_SL_string}, we analyze the spin-orbit contribution from  the `Thomas precession" emerging from a QCD string. All the contributions to the spin-orbit potentials are compared in section \ref{sec_SL_summary}. Our concluding remarks are summarized in section~\ref{sec_CON}. Some additional 
details can be found in the Appendices.

\subsection{Complete and incomplete tunneling in the  QCD vacuum } \label{sec_tunneling}
Instantons 
 describe  
tunneling between gauge field configuration with different Chern-Simons number
\be \Delta N_{CS}\equiv  N_{CS}(\tau=-\infty)- N_{CS}(\tau=\infty)=1
\ee
 Anti-instantons ($\bar I$) describe the reverse tunneling. 
 Since the axial anomaly leads to non-conservation of the $U(1)_A$ quark axial current by the amount
 proportional to the topological charge, or change $\Delta N_{CS}$ of the Chern-Simons number, such transitions  change quark chiralities as well. This is captured by
 the appearance of fermionic zero modes. 
 Therefore instantons -- completed topological transitions -- generate multi-fermion transitions described by the  t' Hooft Lagrangian \cite{tHooft:1976snw}. 

 The earlier works on instantons focused on the spontaneous breaking of chiral $SU(N_f)$ symmetry. Indeed,
in the case of two quark flavors, the t' Hooft Lagrangian  is a 4-fermion operator, similar (but not equal) to the
hypothetical Nambu-Jona-Lasinio local interaction,  originally proposed to explain it. The original
Instanton Liquid Model (ILM) 
\cite{Shuryak:1981fza} focused on
the magnitude of the quark condensate, properties of the pions and other chiral phenomena, for a review see e.g.\cite{Schafer:1996wv}. 

In this paper however,  we address
heavy quarkonia only, and do  not need to consider the fermionic zero mode. We will focus on the  correlations of the gauge fields  along straight paths of heavy quarks.
They are dominated by the (more numerous)
$I\bar I $ pairs or ``molecules", with zero total topology and zero $\Delta N_{CS}$.  Physically, they
correspond to ``unsuccessful tunneling", with a quantum path venturing inside the
topological barrier and returning back. 
(Their separation from the
 small-amplitude fluctuations described by
a quadratic potential and ``gluons" is a matter of convention. This issue is important but will not be discussed further in this work. We delegate it to lattice works, where specific practical definitions are given.)

While single instantons and ``molecules" may have have comparable actions, they have different number of deformations,
and on general ground it is hard to compute their densities. As we noted earlier, 
 the lattice 
studies in~\cite{Athenodorou:2018jwu} have shown that ``molecules"  are more numerous than isolated instantons. 
Their possible contribution in interquark potentials (central and spin-dependent ones),  were explored in 
\cite{Shuryak:2021fsu}. Here, we will follow on this work,
by  calculating and comparing interquark
potentials induced by instantons and strongly correlated $I \bar I$``molecules".

Historically, overlapping $I\bar I$ configurations were first considered in
the framework of the famous quantum mechanical  problem of a quartic  double-well potential. As it was first shown e. g. in \cite{Shuryak:1987tr},  keeping certain points on the path fixed, and minimizing the action, one finds certain {\em conditional minima}. Action minimization along the direction of its gradient 
 (now known as ``gradient flow") were called ``streamline configurations". (The
 corresponding lines are also known in complex analysis as  "Lefschitz thimbles".) 

In gauge theory, the derivation of the gradient flow equation  were pioneered in~\cite{Balitsky:1986qn,Yung:1987zp}. The solution 
was first found 
for large distances (small overlap) of $I\bar I$, and then at all distances in 
\cite{Verbaarschot:1991sq}. It was achieved via a special conformal transformation of $I$ and $\bar I$
into a co-centring configuration. The resulting configurations were also used
in~\cite{Ilgenfritz:1988dh} in finite temperature QCD, and in~\cite{Shuryak:1991pn,Khoze:1991sa} in the semi-classical theory of the so called {\em sphaleron production }. We will continue our discussion of the ``molecular configurarions" in section \ref{sec_molecules}.

(Note that the  sphaleron-like objects are  
still not  observed experimentally, neither in  high energy colliders , nor in QCD or electroweak gauge theory, for a recent discussion of such experiments see e.g. \cite{Shuryak:2021iqu}.)

\subsection{The interquark potentials } \label{sec_potentialsX}
The general theory of heavy quark interactions in QCD was developed by Wilson \cite{Wilson:1976zj},
who introduced the concept of  ``Wilson lines", the path-ordered-exponents of gauge potentials, and related them to  static 
central potentials on the lattice. The theory of QCD spin-dependent forces, as relativistic 
$O(v^2/c^2)$ corrections, was worked out by \cite{Eichten:1980mw}, for subsequent review 
of the non-relativistic QCD reduction see \cite{1707.09647}.

Here we briefly recall (what has already been discussed in \cite{Shuryak:2021fsu}) some aspects about Wilson lines 
$$ W={\rm Pexp}\bigg(i \int A_\mu^a (\tau^a/2) dx^\mu\bigg)$$ As shown in Fig.~\ref{fig_cube}, the location of the Wilson lines
is kept fixed, here going into the $x^4$-direction, crossing the $x^4=const$ plane at $x=(0,0,\pm r/2)$, with the inter-quark distance denoted by  $r$. 

Due to the  ``hedgehog" structure of $A_\mu^a$ in the instantons, the color rotation along the Wilson line
is around a fixed direction. This feature simplifies 
considerably the evaluation of the Wilson line, as
the  path-ordered exponential is traded for an exponential
of an integrated  rotation angle.
All earlier calculations of the instanton-induced potentials, from \cite{Callan:1978ye}
to our work \cite{Shuryak:2021fsu}, used this observation.

The evaluation of  Wilson lines piercing ``molecular" configurations is technically  more involved. The existence of two centers takes away the previous observation, since at different points color rotations are around different vectors. This means that the 
 path-ordered exponents are now incremental products of matrices along the Wilson line. Also, the 4d moduli of the molecular fields is more involved, as detailed in the  Appendix and the sections below. As a result, the
 evaluation of the Wilson loop requires multidimensional simulations using Monte-Carlo.

Before discussing the procedures and results, let us introduce some special cases. 
 If the line between the centers in a molecule $\hat n$ is directed 
 along the (Euclidean) time $x^4$ ( $\hat n=(0,0,0,1) $ or $\chi=0$), the integrals along the Wilson lines  
 vanish, 
 $$\int_{-\infty}^{\infty} A^4 dx^4=0$$
 since $A^4$ is an odd function of $x^4$. It does not happen if $\hat n$  
points in other directions. The integrands are not odd, and the Wilson lines get nonzero values. The correlator of magnetic fields located at two Wilson lines (related to the spin-spin potential) 
 for different orientations of $\hat n$ will be shown below in  Fig.~\ref{fig_three_orientations}.

In this paper we discuss a standard problem, that of the central and spin-dependent interactions of quarks and antiquarks,
in the simplest hadrons, i.e. heavy quarkonia $\bar b b, \bar c c$. In these hadrons the quarks move slowly, and the nonrelativistic
Schroedinger equation can be used, with pertinent potentials. It is well described by a Cornell-like potential, the sum of the perturbative Coulomb $\sim 1/r$ plus confining $\sim \sigma r$ potentials. Below, we will also use the  wave functions generated by this standard procedure. 

In fact two problems appear when one tries to relate this linear potential to the electric flux tube or the  QCD string.   The first  is that the QCD string is quantum not classical, and its contribution is  $\sigma r$
only at large distances. As shown by Arvis~\cite{Arvis:1983fp}, it is even absent at distances $r < \,0.5 \, {\rm fm}$,  that are relevant
to heavy quarkonia. As we noted 
in
 \cite{Shuryak:2021fsu},  an alternative origin of the linear  potential at such distances,  can be provided by instantons. 
 (Below we will show that it is  also the case for an ensemble of ``molecules" as well.)

The second serious 
problem of the flux tube model
is related to the spin-dependent forces. In particular, the spin-spin and tensor
forces  are related to a correlator of two magnetic fields. These are absent 
in gluo-electric  flux tubes. However, instantons   carry magnetic fields. As we will see below, inside the $\bar I I$  configurations, the magnetic fields can even exceed the electric fields.

We start our phenomenological discussion of the spin-dependent potentials, by focusing on the (almost local) spin-spin potential  $V_{SS}(r)$. Four decades ago, the
main source of information about it was the splitting of two basic $1S$
states, $J/\psi$ and $\eta_c$. Today we know also the $2S$ splittings, both in charmonia and bottomonia, 8 states and 4 splittings. Since the sizes of all
these quarkonia are different and their wave functions readily available, 
we can use these set of data to quantify the $range$ of  $V_{SS}(r)$, see
section \ref{sec_SS_pheno}.

The spin-spin $V_{SS}(r)$ is the first spin-dependent potential which has been determined on the lattice
 \cite{Koma:2005nq}. Surprisingly, the numerical results show that it is not local, but yet very short ranged 
\be V_{SS}(r)\sim {\rm exp}(-\beta r), \,\, \beta\approx 3\, {\rm GeV} 
\ee
As we will show below, it is in agreement with all four splittings 
($\bar b b, \bar c c$ at 1S and 2S states) we have studied. 

In section
\ref{sec_SS_molec} we will calculate $V_{SS}(r) $ using  two models, an ensemble of independent
 instantons, and also that of {\em strongly correlated $\bar I I $ molecules}.
 The latter produce stronger potentials, that are approximately exponential, but turn negative at large distances. 
 A combination of this potential with the perturbatively local term ($V_{SS}(r)\sim \delta(\vec r)  $), 
will be shown to reproduce well the four quarkonia splittings. 

Encouraged by this agreement, we take on the least understood 
spin-dependent force, the spin-orbit $V_{SL}$ potential.
In quarkonia its matrix elements can be deduced from splitting 
of the ``opposite parity" $1P$ states $\chi_{c,b}$. This was discussed in the literature many times, including our first paper \cite{Shuryak:2021fsu}.
Such phenomenological analysis  should include necessarily all the three spin-dependent
potentials $V_{SS},V_{SL}$ and the tensor force $V_T$. So, with still limited
phenomenological inputs available, the analysis  points to some problems with $V_{SL}$, at least with
the version suggested by perturbative theory. 

A while ago,  Isgur and Karl \cite{Isgur:1978xj} by considering the 1P-baryon states, arrived at a 
puzzling conclusion. The spin-orbit potential  $V_{SL}$ should be much smaller than suggested by perturbation
theory. In our recent work \cite{Miesch:2023hjt} we re-derived  the wave functions of all five $N^*$ from the 1P-shell, by novel methods based solely on permutation group representation theory. With 5 masses and two mixing angles as input, we  obtained the phenomenological
values of all  matrix elements. Indeed, we observed that the spin-orbit $V_{SL}$ contribution is,
within error, consistent with zero.


\section{Correlated instanton-antiinstanton pairs,\\ or ``molecules"}
\label{sec_molecules}

We already noted that paths describing incomplete tunneling
can be approximated by strongly overlapping $I\bar I$ pairs.
Such
 field configurations are seen in numerical lattice simulations, after just few steps of the
 ``gradient flow" filtering out the gluons. With further gradient flow they pair annhilate, leaving behind a dilute ensemble of  $I$ and $\bar I$ stable under action minimization. It is this ensemble which has been known as the  "instanton liquid model" \cite{Shuryak:1981fza}. An example of  a relatively recent lattice studies in which ``molecules" and their pair annihilation through gradient flow are studied, can be found in~\cite{Athenodorou:2018jwu}.

\subsection{Ratio ansatz for molecules}
\label{sec_ratio}
Failed tunneling configurations can be described by Verbaarschot's or Yung's 
``streamline" field configurations. However, since the conformal mapping used is rather complicated to implement,  we will use a simpler (but sufficiently
accurate) ansatz to describe them, a variant of the so called ``ratio ansatz"
\be \label{eqn_ansatz}
A^{\mu a}(x)={ \bar\eta^{a\mu\nu} y_I^\nu \rho^2/ Y_I^2 + \eta^{a\mu\nu}y_A^\nu*\rho^2 /Y_A^2 \over
1+\rho^2/Y_A +\rho^2/ Y_I }
\ee
where $I,A$ stand for instanton and anti-instanton.
Their centers are  located at $y_{I,A}^m=x^m,m=1,2,3$ and $y^4_I=x^4-R/2, y^4_A=x^4+R/2$, with $Y_{I,A}$  referring to their squared lengths, e.g.
$Y_A=(y_A^\mu y_A^\mu)$. Near one of the centers (e.g. $Y_I$ small and $Y_A$ large),
the dominant contribution in the numerator is the first term, and  with the leading terms in  the denominator, yield the familar field of an instanton in singular gauge. 

\begin{figure}
    \centering
    \includegraphics[width=7cm]{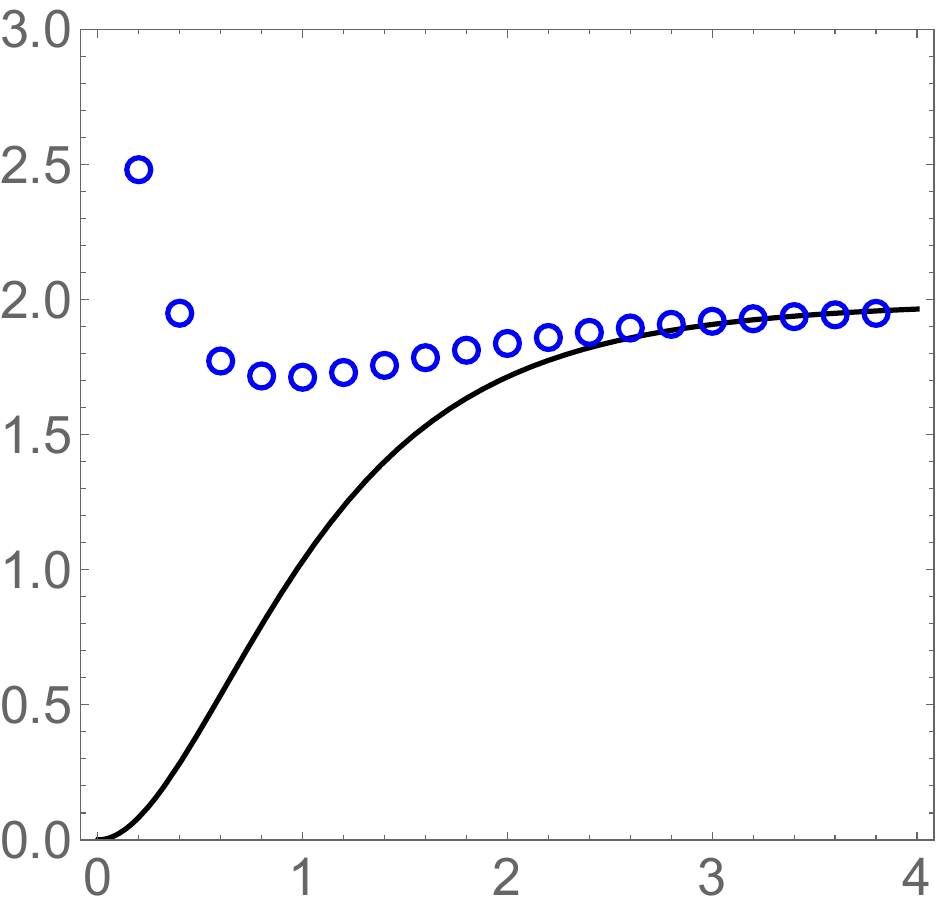}
    \caption{The action of the $I\bar I $ configurations (in units of single instanton action $8\pi^2$) as a function of distance between centers $R/\rho$.
    The solid line show that for the ``streamline configuration" \cite{Verbaarschot:1991sq}, the points are for our ansatz (\ref{eqn_ansatz}). 
    }
    \label{fig_action}
\end{figure}

\begin{figure}
    \centering
    \includegraphics[width=7cm]{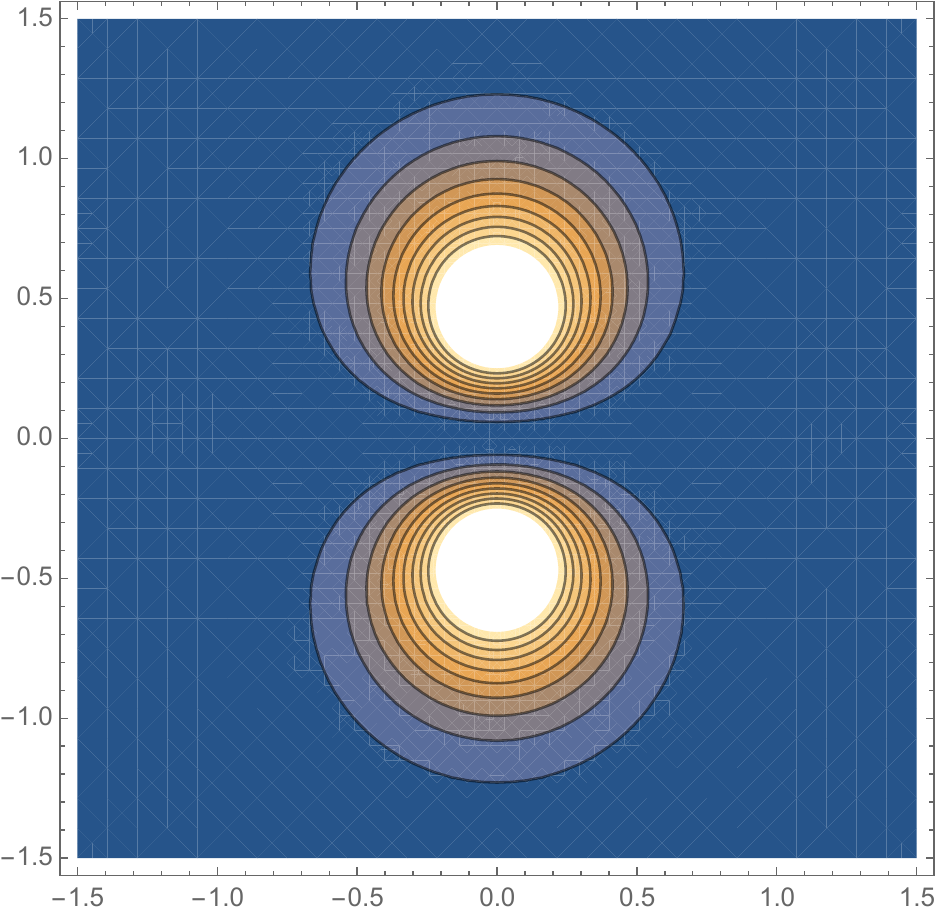}
    \caption{Contour plot of the action density as a function of $x^1,x^4$ coordinates,
    in units of instanton size $\rho$. The distance between the centers $R=\rho$, and $R^\mu$ is directed along the Euclidean time axes $\tau=x^4$.}
    \label{fig_molec_R1}
\end{figure}

It is a simple  task for Mathematica to produce  workable 
 expressions for the field strength and their squares (yet still too long to be given here). The distribution
of the action density $[G_{\mu\nu}^a]^2$ 
is shown in Fig.\ref{fig_molec_R1}. The integrated action (in units of the instanton action $S_{0}=8\pi^2/g^2$) at large $R\rightarrow \infty$ is twice that of the instanton. However, at small $R<0.5$ it does $not$ go to zero (as the streamline configuration does) but display a
small repulsive core. That does not affect the calculations, since 
as a representative of the vacuum  molecules, we will use  $R=\rho$.
There are small deviations between the streamline and the proposed ansatz (\ref{eqn_ansatz}), but
they are numerically small.

\begin{figure}
    \centering
    \includegraphics[width=7cm]{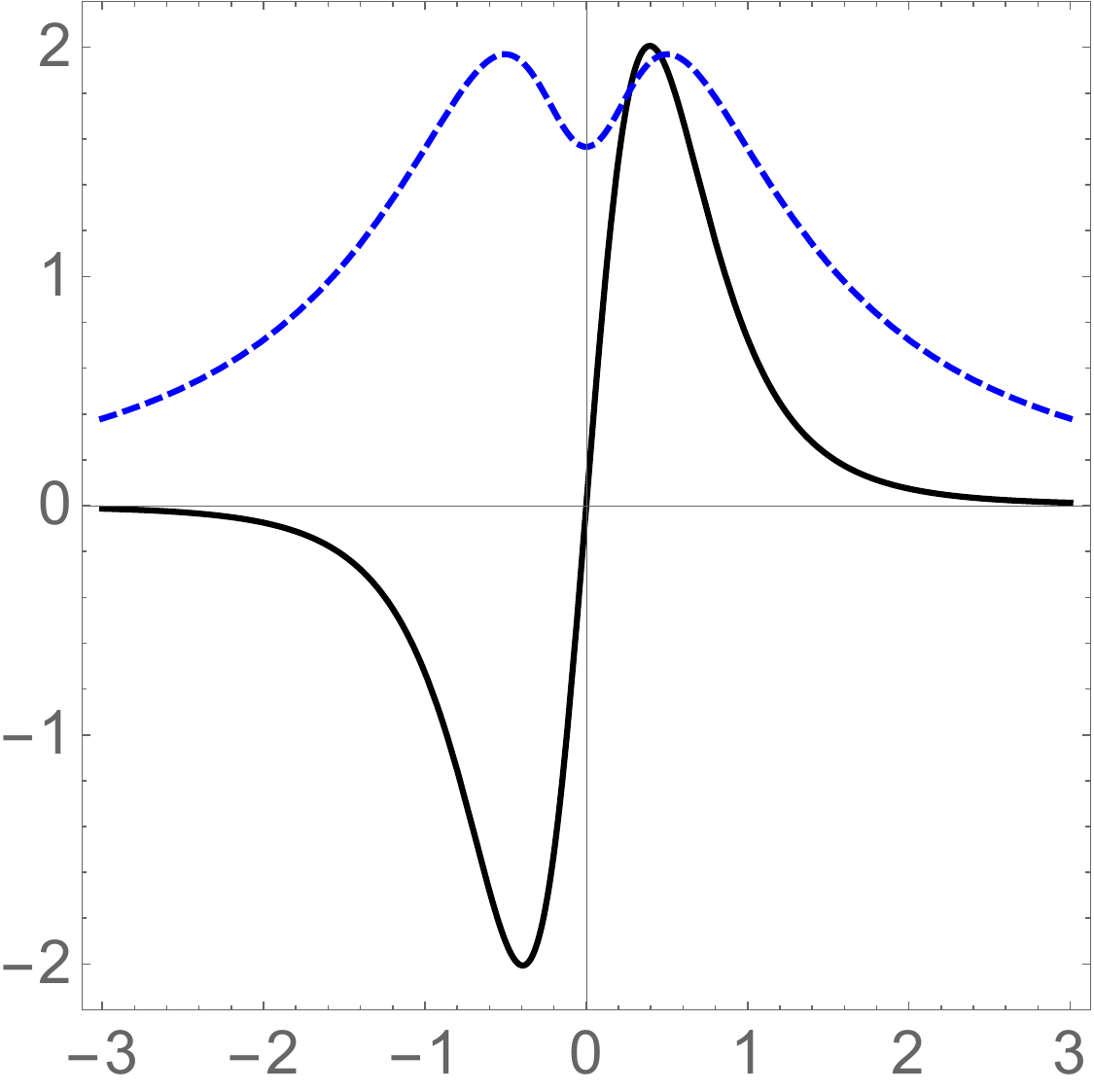}
    \caption{The electric (black-solid line) and the magnetic (blue-dashed line) field components $\mu=3,a=3$, as a function of $x^4/\rho$. }
    \label{fig_E_B_33}
\end{figure}

The behavior of the gauge fields  are shown in Fig.\ref{fig_E_B_33}.
Note that the electric field is odd in $x^4\rightarrow -x^4$, while the magnetic field is even. Also $A^4$ is odd,
and therefore the configuration  at the $x^4=0$ location in the 3D subspace is purely magnetic,  the so called ``sphaleron path". The molecular configurations therefore can be
interpreted as tunneling amplitude and anti-amplitude, separated  by the analogue  
of quantum-mechanical ``turning point", at which $\vec E$ (the momentum) vanishes. 

\subsection{Central potential from uncorrelated instantons and molecules} 
\label{sec_potentials}
A conjecture that the central inter-quark potentials at intermediate distances ($r\sim 1/2-1 \, fm $) can be due to  a ``dense instanton 
ensemble" was proposed in our paper \cite{Shuryak:2021fsu}. The calculations have been carried for uncorrelated instantons, with the only thing changed being their density.
While some qualitative discussion of changes due to $I\bar I$ molecules have also been made,  in particular the dependence on their orientations,  we have not actually calculated  the Wilson lines for correlated $\bar I$ pairs, due to technical difficulties.

\begin{figure}[h!]
    \centering
    \includegraphics[width=8cm]{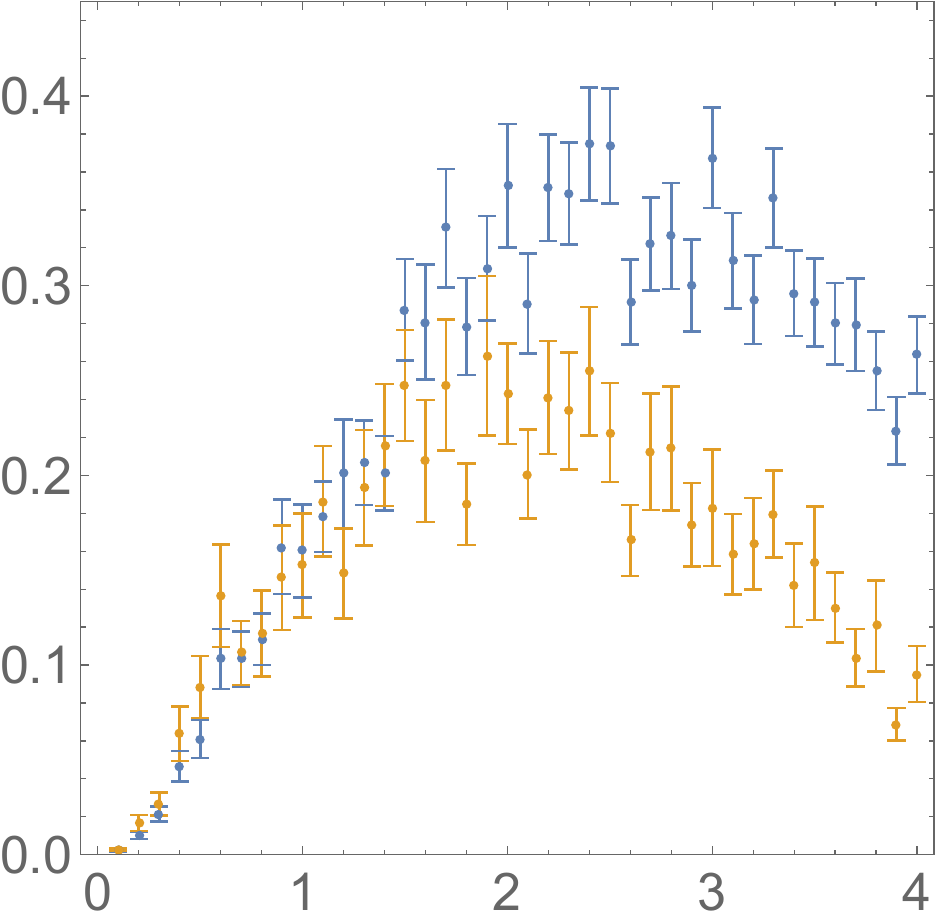}
    \caption{Plot of $Tr(\hat 1 - W(-r/2)W^+(r/2)) $ as a function of the distance   $r/\rho$ between temporal Wilson lines.
    The blue (upper on the right) points are for a single instanton, and the red (lower) points are for an $I\bar I$ molecule.}
    \label{fig_WW_plot}
\end{figure}

We now perform this calculation, much like 
it is carried on the lattice. The Wilson lines are divided into segments, typically with size fraction of $\rho$, for which the exponent of the local field is calculated, with subsequent multiplication of resulting unitary matrices along the Wilson loop.  A comparison of ``single instanton" and "molecular" results is shown in Fig.\ref{fig_WW_plot}. The quantity plotted times the scaled density factor $n_{I,\bar I I}\rho^3$ is the static central potential. 

The main feature stemming from this plot is that
in the range  $r\in [0,2\rho]=[0, 0.66\, fm] $,  the potential is linear, with {\em the same slope} for both single and ``paired" instantons. It means that, with proper density, both ensembles
can generate linear nonperturbative potential. Note that
this is precisely the intermediate range for which the string
interpretation is problematic (which is manifest  in Arvis resummed central potential). 

At larger distances $r>2\rho$ `` the molecules" produce a smaller potential  than uncorrelated instantons. This happens because $A_0$ and the electric field are odd (sign changing) along the line connecting
the two centers of the molecule, so the Wilson line rotation angles along these directions cancel out. 

In our recent paper \cite{Miesch:2023hjt}, we extended these studies to the  ``hypercentral" static potentials induced by instantons in
$ccc$ baryons and $cc \bar c \bar c$ tetraquarks. While in the former we found it to be due to a sum of binary potentials, it is not the case for the latter. While doing the calculations
for this paper, we also evaluated the potentials for the baryons induced
by ``molecules". We found that it has the same shape as for uncorrelated instantons.

\section{Spin-spin potentials} \label{sec_spin-spin}
\subsection{ Spin-spin forces in S-shell quarkonia: phenomenology }
\label{sec_SS_pheno}
The perturbative SS forces (as is well known in atomic and nuclear physics)
are related to a Laplacian of the Coulomb potential, and thus have zero range
(delta function). The nonperturbative forces, on the other hand, are related with finite-size configurations,  and therefore must have a finite range. 
What is the value of this range we will evaluate $phenomenologically$, from data, then compare it to $V_{SS}$ obtained in lattice simulations, and eventually to what we have calculated
from our model of the QCD vacuum.

We start with the naive point of view, by extending a massless gluon exchange to a massive one with an effective mass $M$. The result, is 
Yukawa-like exchange potential  $exp(-M \cdot r )$. The lowest glueballs have masses $\sim 2 \, GeV$, so perhaps $M\sim 1 \, GeV$.

 Let us now have a look at the available lattice results \cite{Koma:2005nq} for
 charmonium $V_{SS}(r)$. The simplest parameterization of their numerical data take the exponential form
\be \label{eqn_VSS_lat}
V_{SS}= \alpha\,e^{-\beta r}, \qquad (\alpha, \beta)=(2.15, 2.93) \,{\rm GeV} 
\ee
Note that the range quoted for the  parameter $ \beta^{-1}\approx (1/15)\, fm$ is surprisingly short, three times shorter than the naive argument above. It may not be very accurate as it is close to the UV scale of the simulation (the lattice spacing), but it still shows  that 
in the QCD vacuum the magnetic fields are not only strong but also {\em extremely inhomogeneous}.

\begin{figure}
    \centering
    \includegraphics[width=8cm]{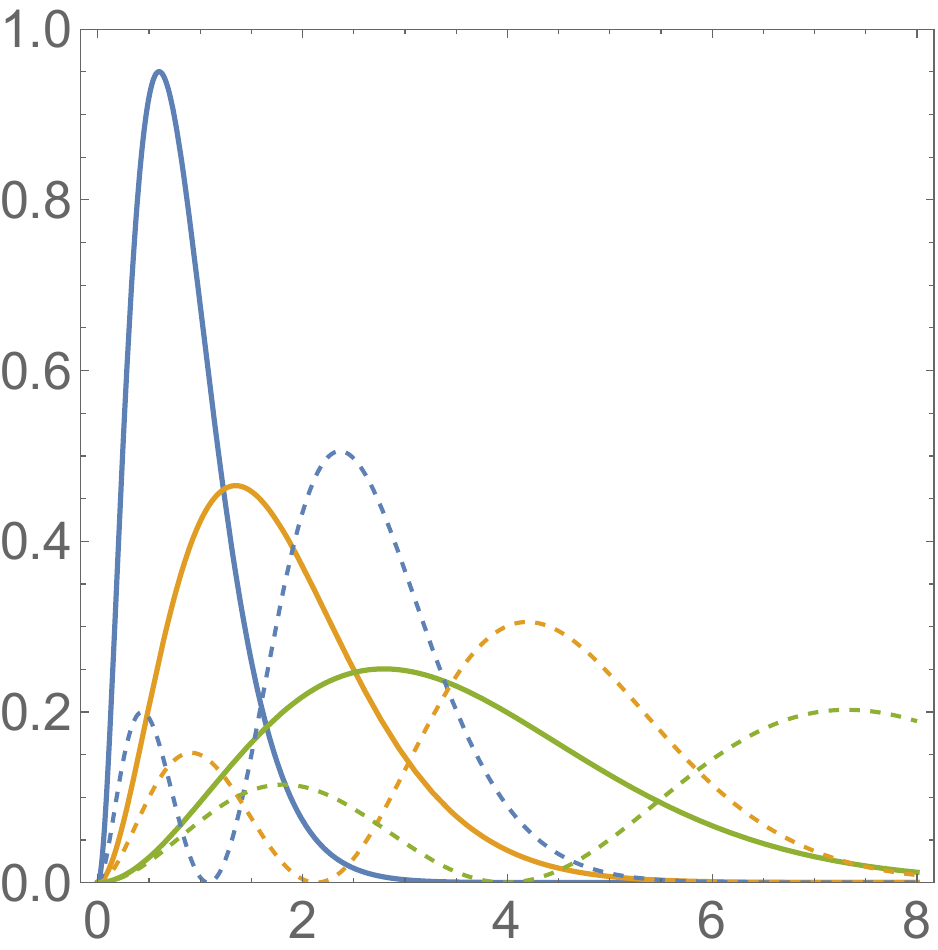} 
    \caption{Wave functions $|\psi(r)|^2 \cdot r^2$ of 1S (solid lines) and 2S (dashed) of quarkonia $b\bar b, c\bar c, s \bar s$ mesons (peaked from left to right) versus the distance $r\,\,\ (GeV^{-1})$. One $fm$ is approximately $ r\approx 5 \,(GeV^{-1} $.}
    \label{fig_wfs}
\end{figure}

Is this observation supported by experimental data? To answer this question in principle, we need to access the matrix elements of this potential  for a set of hadrons of different sizes. In fact, we know rather well the wave functions of quarkonia, and as shown in
Fig.\ref{fig_wfs} they do cover a rather large range of sizes. 

We will now use eight quarkonia states, or
four known splittings of the lowest S-shell states (L=0),  
between  the spin $S=J=1$ states and $S=J=1 =0$, in $\bar b b,\bar c c$ quarkonia  
\ba 
\Delta_c^{S1} &\equiv& M_{J/\psi}-M_{\eta_c}\approx 116 \, MeV \nonumber \\
\Delta_c^{S2}  &\equiv& M_{2S\psi }-M_{2S\eta_c}\approx 51 \, MeV \nonumber \\
\Delta_b^{S1}  &\equiv& M_{\Upsilon}-M_{\eta_b}\approx 61\, MeV \nonumber \\
\Delta_b^{S2}  &\equiv& M_{\Upsilon2S}-M_{\eta_b2S}\approx 24\, MeV 
\ea
(Since $\eta_b^{2S}$ is not yet firmly established according to PDG,
the last quote may not be quite accurate.)

Since we are interested in the range of $V_{SS}$, we focus first on the three ratios 
\ba
\label{RATIOX}
R_c^{S2} &\equiv& {\Delta_c^{S2}\over \Delta_c^{S1}}=  0.44\nonumber \\
R_b^{S1} &\equiv& {\Delta_b^{S1} \over \Delta_c^{S1}}= 0.53 \nonumber \\
R_b^{S2} &\equiv&  {\Delta_b^{S2}\over \Delta_c^{S1}}=  0.20
\ea
If the SS potential has zero range, i.e. $V_{SS}\sim\delta^3(\vec r)$, as suugested by 
one-gluon exchange, the ratios (\ref{RATIOX}) should  be given by the ratios
of the wave functions at zero distance. Using the wave functions already evaluated, we find that zero range leads to 
\ba R_c^{S2}&=&|{\psi_{cS2}(0) \over \psi_{cS1}(0)}|^2\approx \nonumber 0.52 \nonumber  \\
R_b^{S1} &=&|{\psi_{bS1}(0) \over \psi_{cS1}(0)}|^2 ({M_c \over M_b})^2\approx  1.58 \nonumber \\
R_b^{S2}&=&|{\psi_{bS2} (0)\over \psi_{cS1}(0)}|^2 ({M_c \over M_b})^2\approx 0.18
\ea
where for bottomonium we rescaled $V_{SS}$ by the squared quark masses. The agreement with 
the empirical ratios \ref{RATIOX}),
is quite poor, especially for $\Upsilon / J\psi$. Some effect can be due
to the running $\alpha_s$, but it can hardly give a factor of 3 needed.

\begin{figure}
    \centering
    \includegraphics[width=6cm]{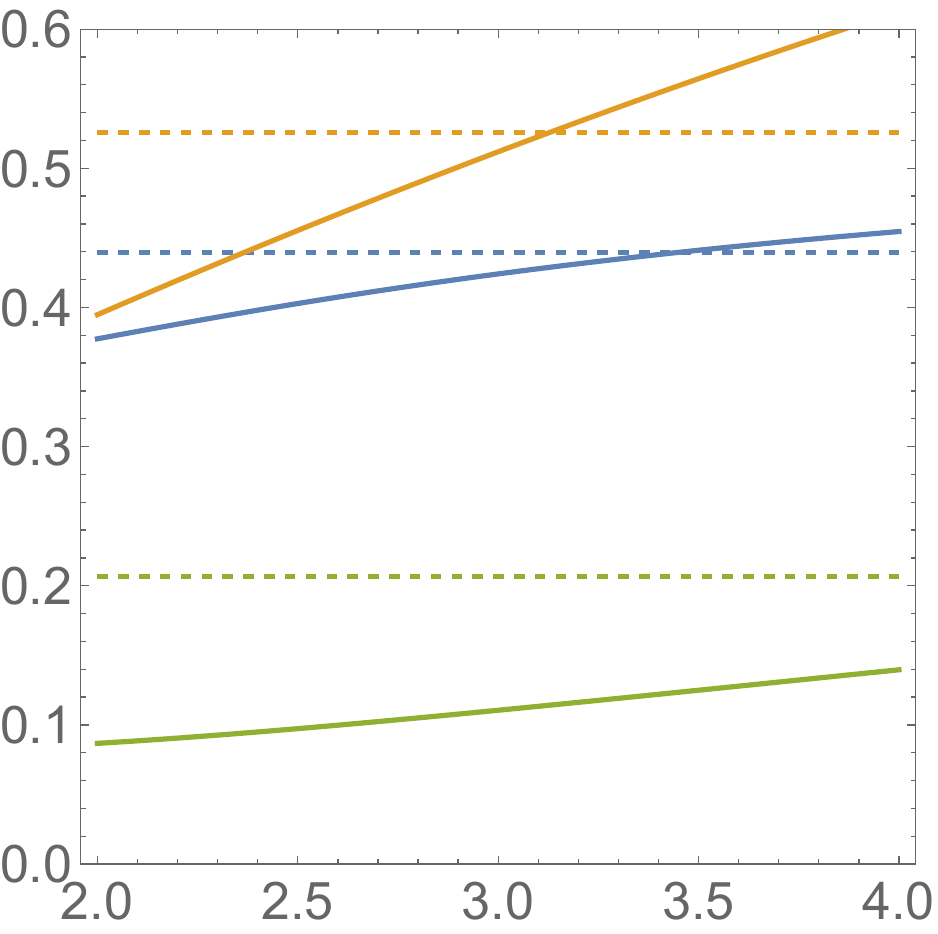} 
    \caption{The horisontal dashed line indicate values of the three experimental ratios $R_b^{S1},R_c^{S2},R_b^{S2}$, top to bottom. Their theoretical values are
    shown by solid lines as a function of range parameter $\beta (GeV)$}
    \label{fig_SS_ratios}
\end{figure}

With an  exponential parameterization (\ref{eqn_VSS_lat}) for the range, 
we first observe that in the ratios the parameter $\alpha$ drops out, with a dependence only on $\beta$. In Fig.\ref{fig_SS_ratios} we show 
these ratios versus $\beta$. The two well known experimental ratios can be accurately  reproduced with the values $\beta_{fit} \sim 3-3.5 \, GeV$,
which is quite close to the quoted lattice value. (The accuracy of the last ratio is suspect.)  Moving from the ratios to 
the  absolute magnitude of the four matrix elements,  we find that if $\alpha$ is 
 increased by $30\%$, all matrix elements are 
 well described.
 
Conclusion: The lattice study giving $V_{SS}$ as in (\ref{eqn_VSS_lat}) is consistent with
 current phenomenology. This is rather impressive since the lattice work only used the $1S \,\bar c c $ state for comparison,  with  none of  the three other states considered.

\subsection{Spin-spin forces induced by instanton-antiinstanton molecules } \label{sec_SS_molec}
The spin-spin and tensor forces originate from the interaction of the quark magnetic moments with the vacuum  magnetic fields. In \cite{Eichten:1980mw} they are related to the potential $V_4$ 
by the following correlator 
\be \label{eqn_BB}
V_4(r)\equiv \int_{-T/2}^{T/2}\int_{-T/2}^{T/2}
{d\tau d\tau' \over T} \langle W B(\vec z,\tau) W B(\vec z',\tau') \rangle  
\ee
where $r=|\vec z-\vec z'| $ and $W$ symbolically refer to the  Wilson lines connecting the
two magnetic fields around the original rectangular contour. 

In our paper \cite{Shuryak:2021fsu} we have evaluated  $V_{SS}$ induced by the
instantons, and found that its matrix elements are comparable to the perturbative ones.  Taken together, they approximately reproduce
the spin-spin splittings in charmonium. Yet its range failed to be as short
as the phenomenology and lattice suggest.

\begin{figure}
    \centering
    \includegraphics[width=6cm]{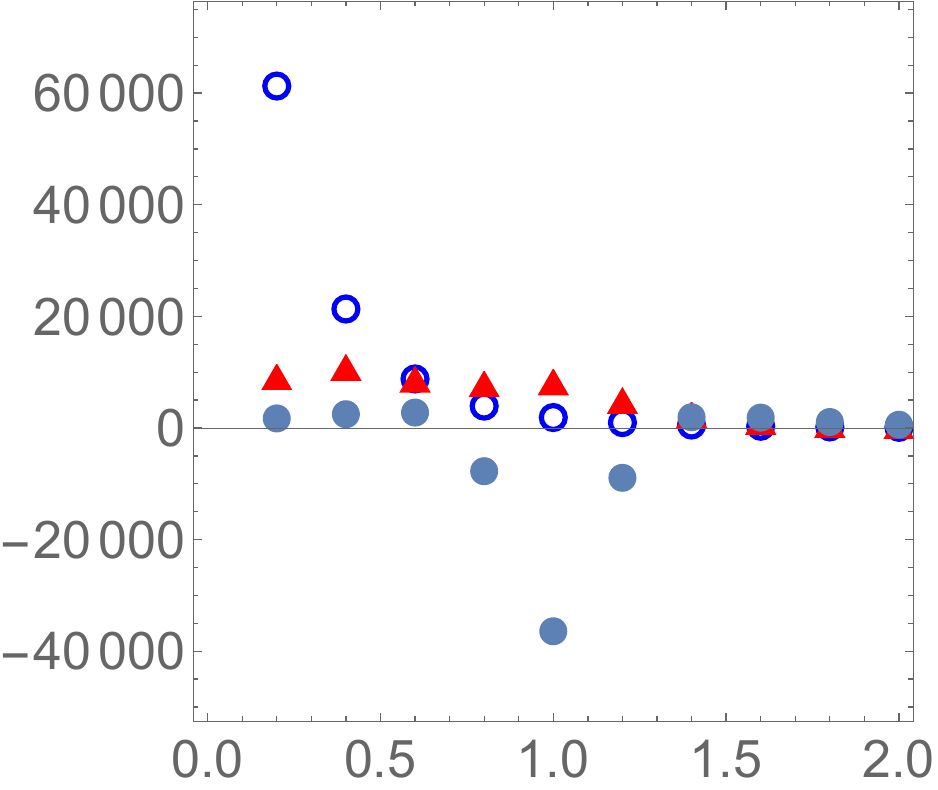}
    \caption{The BB correlator as a function of interquark distance, in  $r/\rho$ units. The blue open points, the red triangles and the closed points correspond to 
    the $I\bar I$ rotated by an angle $\alpha=0, \pi/4,\pi/2$, respectively.
    In this plot there is no shift of the configuration, only a rotation in
    the 3-4 plane.
    }
    \label{fig_three_orientations}
\end{figure}

In the appendix in~\cite{Shuryak:2021fsu} we argued that the
shorter range should be given by a ``molecular vacuum", with no  explicit calculation. 
Before we do so below, 
let us qualitatively explain how the BB correlator depends on the space-time orientation
of the line connecting the $I\bar I$ centers. In Fig.~\ref{fig_three_orientations}
we have shown three sets of points, corresponding to different orientation angle
$\alpha$ relative to the (Euclidean)  time $\tau$. A very-short range
contribution is seen for open points, corresponding to  the $I\bar I$ line directed in the time direction. In this case the Wilson lines may pass close to both $I$ and $\bar I$
centers. As the inter-quark distance increases, this is no longer possible and 
the correlator rapidly decreases. If the molecules are rotated by  $45^o$ the maximum
is shifted away from the zero inter-quark distance. If they are rotated by  $90^o$ (closed points)
the maximum is at $r/\rho=1$ and the correlator turns negative, as the fields have opposite directions.

The plot in~(\ref{fig_three_orientations}) is given  for a qualitative orientation 
of the reader only. The  calculation should include two effects, which we now detail.
First, the fields of the $I\bar I$ molecules should be rotated by random $SO(4)$ rotations. In appendix \ref{sec_4d_rotations} we recall how they can  be constructed via two
$O(3)$ angular momenta, each of which with  three Euler angles, so that the random rotations
are defined by 6 parameters. The random generation of ($SO(4)$) orientations is
 explained in section \ref{sec_4d_rotations}, and their use
explained in Appendix \ref{sec_evaluation}.
Second, its center should be shifted by certain random 3d displacement vector, which
we randomly generated in a 3d cube. 

\begin{figure}
     \centering
     \includegraphics[width=8cm]{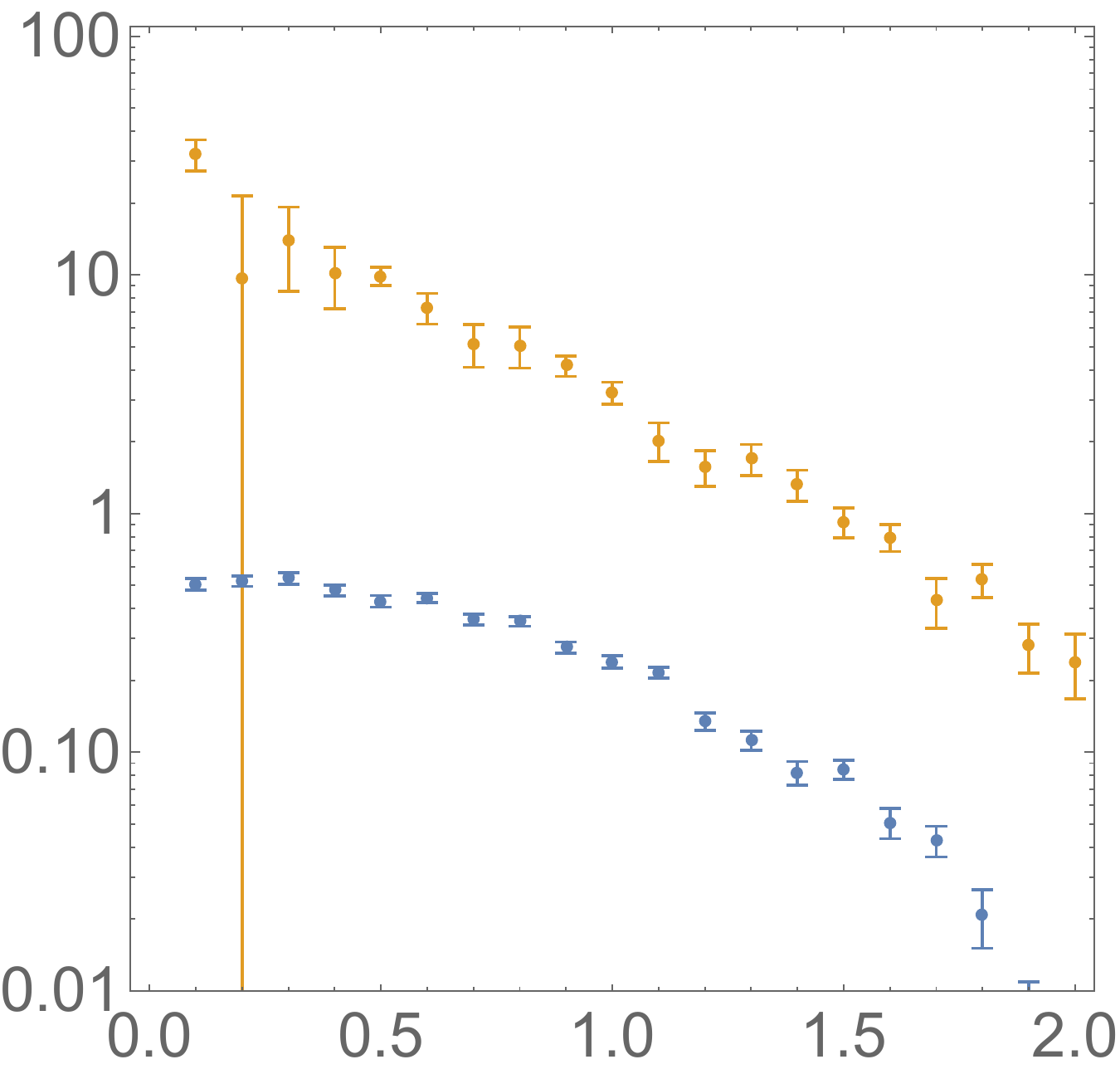}
         \includegraphics[width=8cm]{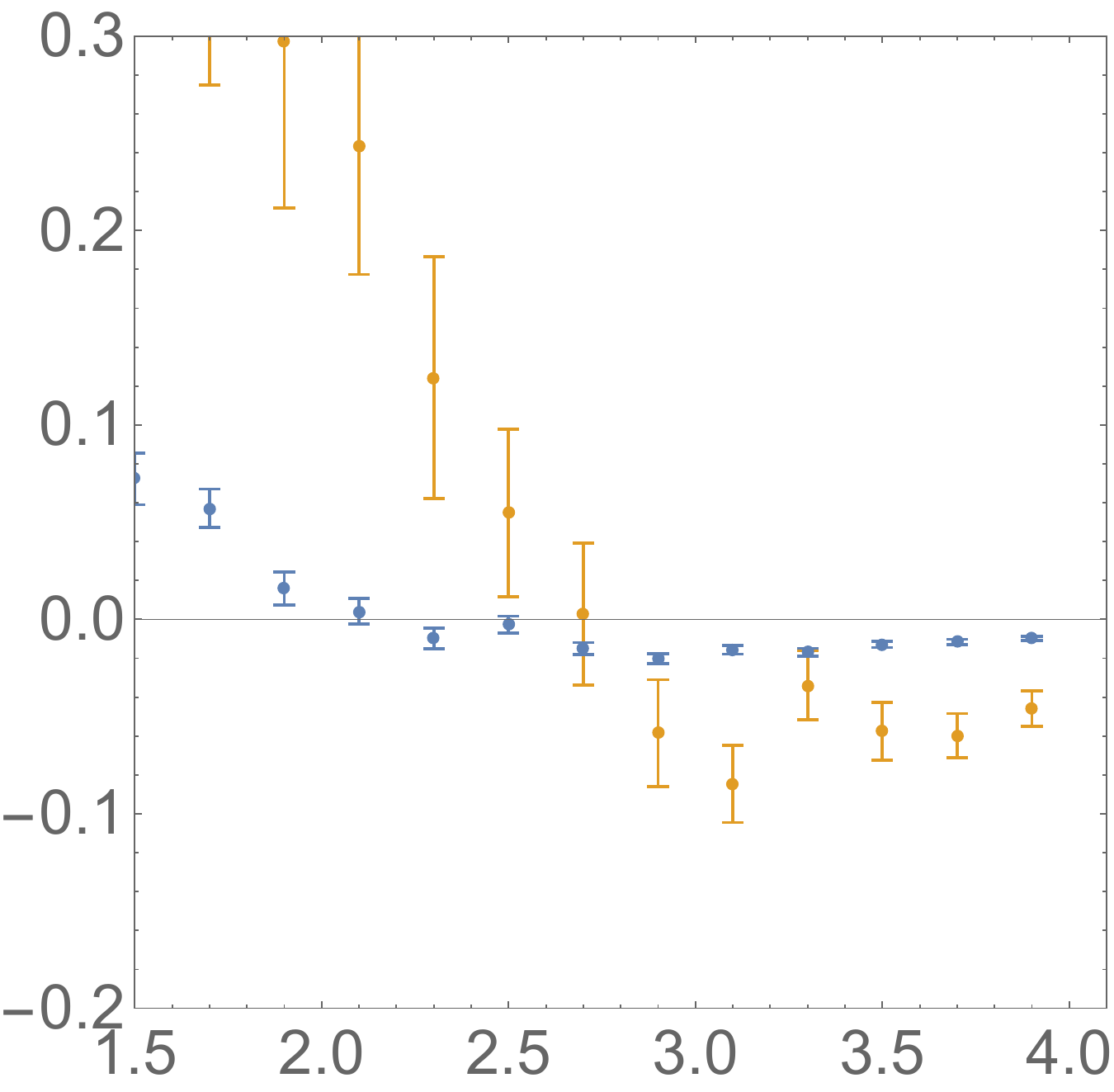}
     \caption{The correlator of magnetic fields on Wilson lines (\ref{eqn_BB}) versus the interquark distance in units of the instanton size, $R/\rho$. Both plots are logarithmic, with the lower plot displaying the negative contributions. The red (mostly upper) points are for $I\bar I$ 
     molecules, and the blue (mostly lower) points are for single  instantons.
     }
     \label{fig_BB_mol_inst}
 \end{figure}

The resulting Monte-Carlo calculation is therefore a sampling in 6+3=9 dimensions.
In addition, there are two integrals over times $\tau,\tau'$  where
the magnetic fields are located.  Since the integrand is rather strongly localized in all these dimensions, proper sampling by Monte-Carlo is hard. The  results are plotted 
in Fig. \ref{fig_BB_mol_inst}, for single instantons or single $I\bar I$ 
molecules. Note that the error bars on the  plot  are 
statistical, calculated from the spread of the points.

The first thing we note from this plot, is that the results for single instantons (blue, mostly lower) and ``molecules" (red, mostly upper)
points are quite different, both in  shape and magnitude.  For $R<2\rho$ the  ``molecules" produce a nearly perfect exponential shape,   while the single instantons produce a nearly  Gaussian shape.  The slope of this exponent is however not as steep as $\beta \approx 3\, GeV$ discussed  above. 

For $R/\rho>2.5 $  both correlators turn negative. The reason  is the  ``hedgehog" structure of the magnetic fields. For close points, the 
$\vec B$ fields  have similar directions, 
but for points on opposite sides from the instanton/molecule center,
the $\vec B$ fields point away from each other. The negative contributions amount to exchanges of mesons of different sizes in   $V_{SS}(R)$.  

The magnitude is quite different as well. Naively, since the  ``molecules"
are made of $two$ objects, one might expect enhancement to be $\sim 2$,
while the Monte-Carlo integration (performed with the same
rotations and shifts for both) indicate instead the enhancement $\sim 10$.
This comes because the magnetic fields of both interfere constructively
in between the two centers.

\begin{figure}[h!]
    \centering
    \includegraphics[width=6cm]{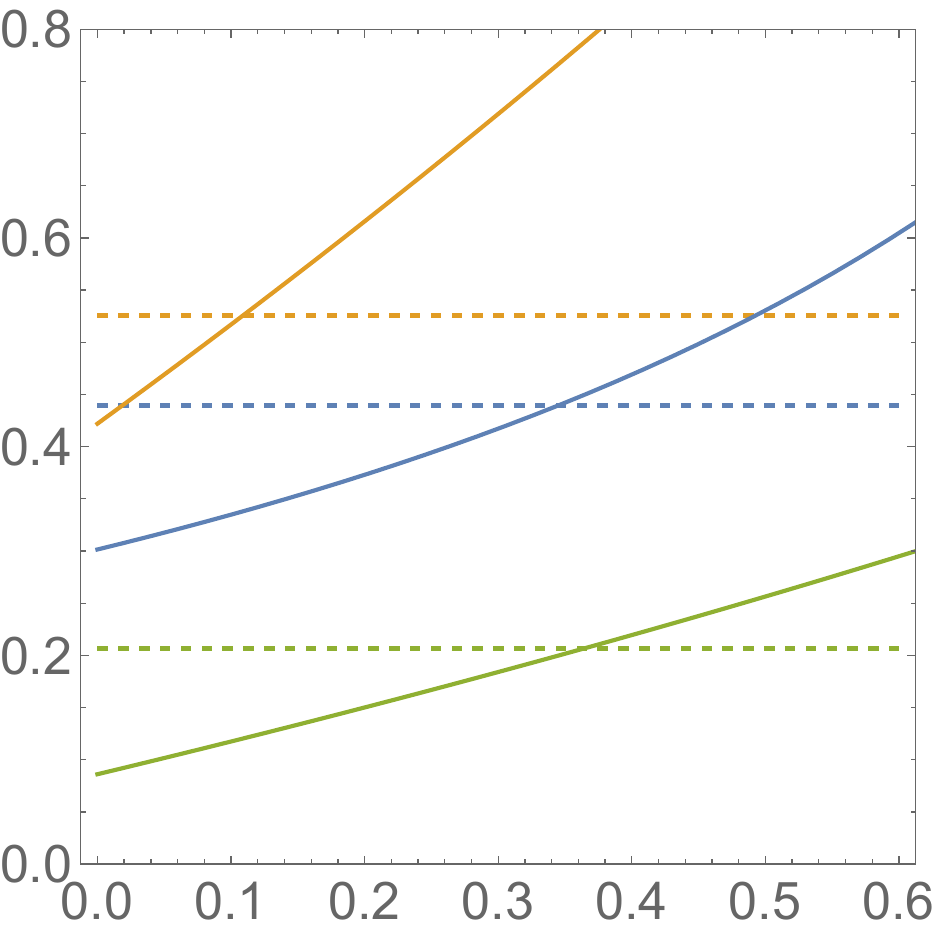} 
    \caption{The horisontal dashed lines refer to the  values of the three experimental ratios $R_b^{S1},R_c^{S2},R_b^{S2}$, top to bottom. Their theoretical values are
    shown by solid lines as a function of  the parameter $\xi$ defined in (\ref{eqn_xi}).}
    \label{fig_xi}
\end{figure}

Note that in the matrix elements of the potential $V_{SS}$, the volume integral is enhanced by $r^2$, 
\be \langle V_{SS}\rangle=\int dr r^2 V_{SS}(r) |\psi(r)|^2 \ee
which enhances large distances relative to short distances.
Furthermore, one should not forget that instantons are not the only contribution.
There is still a  perturbative contribution to $V_{SS} $ at smaller distances.
We introduce a parameter $\xi$, and define all matrix elements as
a superposition of the nonperturbative potentials we derived, plus a local perturbative 
contribution
\be \label{eqn_xi}
\langle V_{SS}\rangle=(1-\xi)\int dr r^2 V_{SS}^{nonpert} |\psi(r)|^2+
\xi |\psi(0)|^2\ee
We evaluated (\ref{eqn_xi}) for four spin-spin splittings, in $\bar c c,\bar b b$ and $1S,2S$ wave functions. In order to postpone the discussion
of the absolute magnitudes, we  focus on three ratios $R_b^{S1},R_c^{S2},R_b^{S2}$, as we did earlier.
In Fig.\ref{fig_xi} we compare the results of the convolution of
$V_{SS}^{mol}(r)$ with 4 respective wave functions, to
their empirical  values 
(horizontal dashed lines). We see that at $\xi \sim 0.3$ all three
ratios are approximately reproduced.  We  conclude that the local perturbative contribution is still essential, albeit not leading in contribution. 

We close  this section with the following comments. We discussed
 two extreme descriptions
of the instanton ensemble in the QCD vacuum. One is an ideal gas
of  uncorrelated  instantons with random spacial/color
orientations. What we call "$\bar I I$ molecule" is, on the contrary, maximally
correlated pairs $\bar I I$ sharing the same color orientation.

In general, a pair $\bar I$ and  $I$
 can have arbitrary relative orientations (which would
correspond to adding $different$  $SU(3)$ color rotation
matrices $\hat U_I$ and $\hat U_{\bar I}$ to the first and second
terms in the numerator of (\ref{eqn_ansatz})). The explicit formulae can be found in~\cite{Verbaarschot:1991sq}. The
``molecules" we used
have minimal action (maximal attraction).  {\em Partially correlated} pairs in instanton-antiinstanton ensembles may produce
results somewhere in between these two limits. This issue was 
studied in statistical simulation of the ensemble, and can be studied more. Also the distribution over
the relative orientations can be  extracted from lattice configurations.

Summarizing, the most important result of the present calculation
is that the ``molecular" contribution to the $BB$ correlator is significantly
larger than that of the ``single instantons". It does have approximately
the exponential dependence on the distance, and, with some admixture
of the perturbative local force, can reproduce all available data
of the spin splittings in charmonia and bottomonia.

\section{Spin-dependent potentials in the P-shell  (L=1) quarkonia} \label{sec_spin_Pshell}
\subsection{Matrix elements of three spin-dependent interactions in the P-shell quarkonia}
\label{sec_pheno}
From the early days of quarkonium spectroscopy, 
it is known that the perturbative  spin-orbit potential  $V_{SL}(r)$
fails to describe the 1P (``wrong parity") 
mesons and baryons. We are unaware of a successful description  of
the non-perturbative part, or any related lattice studies.

We already discussed this topic in our recent works. 
In \cite{Shuryak:2021fsu} we discussed the $1P$ states of mesons, 
from heavy quarkonia to light quarks, focusing on the overall 
dependence on the quark mass. In \cite{Miesch:2023hvl} we focused on five 
negative parity nucleon excitations, also in the $1P$ shell.
While that paper focused mostly on technical issues (such as
the representation of the $S_3$ permutation group, and the explicit derivation 
the wave functions for these states), it also showed that the phenomenological
values of the matrix element $\langle 1P | V_{SL}(r) | 1P \rangle$ are small and consistent within errors, with those originally found by Isgur and Karl \cite{Isgur:1978xj}.

\begin{figure}[h!]
    \centering
    \includegraphics[width=8cm]{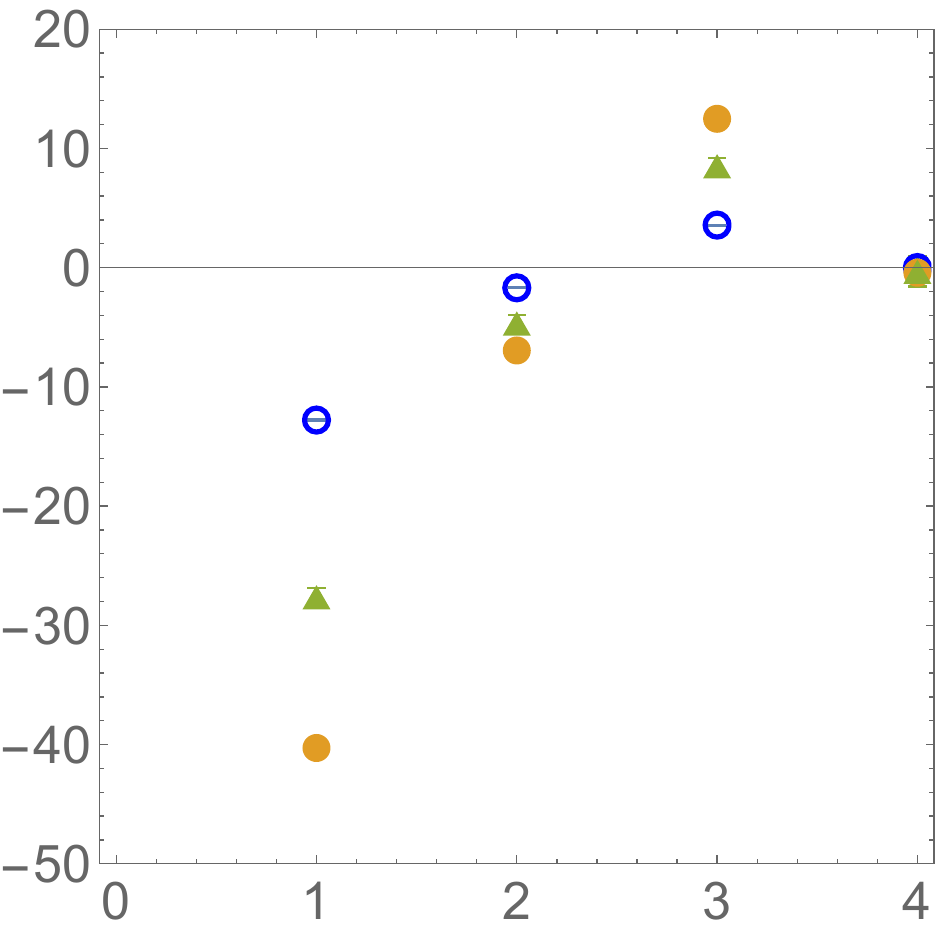}
    \caption{Mass splittings (MeV) from the `spin-unpolarized" mass combinations (\ref{eqn_unpolarized}) of $\chi_0,\chi_1,\chi_2,h$, left to right columns. The three sets of points are for $\bar c c, P1$
    (blue circles), $\bar b b, P1$ (red closed circles) and $\bar b b, P2$ (green triangles). For a better comparison with others, the $\bar c c, P1$
    data are re-scaled by a factor of $(m_c/m_b)^2$.    }
    \label{fig_12_splittings}
\end{figure}

The phenomenological introductory material in the current work
is based on (now well known) $1P$ and $2P$ states in $\bar b b$ and  $1P$ $\bar c c$. Each has four (or $quartet$) states called $\chi_0,\chi_1,\chi_2,h$. Their splittings from the ``spin-unpolarized" mass combinations (\ref{eqn_unpolarized}) due to the spin-dependent forces, are shown in Fig.\ref{fig_12_splittings} and recorded  in  Table \ref{tab_split_compare}. Note that the structure of the splittings for all three quartets are very similar. Note also that the splitting for the $h$ states is much smaller than for $\chi_J$. Since it has total spin $ S=0$,
it is only due to the spin-spin interactions, an indication that its matrix elements is small, for all quartet considered.

The three well known spin-dependent structures are defined by 
\ba V_{SD}&=&\big[\vec S \cdot \vec L\big] V_{SL} \nonumber \\ 
&+& \big[3 (\vec S_1 \cdot \hat \vec r)(\vec S_2 \cdot \hat \vec r)-(\vec S_1 \cdot \vec S_2)\big] V_T \nonumber \\ 
&+& \big[\vec S_1 \cdot \vec S_2 \big]V_{SS}  
\ea
In Appendix \ref{sec_Pshell} we give the explicit spin-orbital wave unctions  of the P1 quartets. Using standard methods (or these wave functions), 
the values of the spin-dependent operators yield
\ba
\langle \chi_0 | V_{SD}| \chi_0 \rangle &=& 
-2\langle V_{SL} \rangle -  \langle V_{T} \rangle+ {1\over 4} \langle V_{SS} \rangle \nonumber \\ 
\langle \chi_1 | V_{SD}| \chi_1 \rangle &=&   -1\langle V_{SL} \rangle -  {1 \over 2}\langle V_{T} \rangle+ {1\over 4} \langle V_{SS} \rangle \nonumber   \nonumber \\ 
\langle \chi_2 | V_{SD}| \chi_2 \rangle &=&  +1\langle V_{SL} \rangle - {1 \over 10} \langle V_{T} \rangle+ {1\over 4} \langle V_{SS} \rangle \nonumber     \nonumber \\ 
\langle h | V_{SD}| h \rangle &=& - {3\over 4} \langle V_{SS}\rangle
\ea

Using the experimental masses (and their error bars), we minimized the standard 
$\chi^2$ for each quartet between data and the first-order expressions 
above. The best-fit values of the matrix elements of the three spin-dependent potentials for each states in the quartet, are given in   Table \ref{tab_fitted_averages}.
The first-order account for the spin-dependent forces, gives a good description of the observed splittings, see Table \ref{tab_split_compare}. 

\begin{table}
    \centering
    \begin{tabular}{cccc} \hline
         & $\langle V_{SL} \rangle$ & $\langle V_{T} \rangle$ & $\langle V_{SS} \rangle$\\ \hline
   $P1 \,\,\bar b b  $     &  13.6  & 9.9  & 2.4 \\
   $ P2 \,\,\bar b b$      &   9.3    & 8.8 & 1.6 \\
   $ P1 \,\,\bar c c  $    & 35. & 40.1 & 1.2 \\ \hline
    \end{tabular}
    \caption{Best fit values for the matrix elements of the spin-orbit, tensor and spin-spin potentials }
    \label{tab_fitted_averages}
\end{table}

\begin{table}[h!] \begin{center} \begin{tabular}{c c c c c }
\hline
&                 $\chi_2$ & $\chi_1$ &  $\chi_0$ & h   \\
\hline
$\bar b b$  P1 & 12.5 $\pm$ 0.5 & -6.9 $\pm$ 0.5 & -40.3 $\pm$ 0.6 & -0.4 $\pm$ 0.8 \\
fit & 12.47 &  -6.91 &-40.31  & -0.44\\
$\bar b b$  P2 &  8.5 $\pm$ 0.7  & -4.7 $\pm$ 0.7 & -27.6 $\pm$ 0.8 & -0.3$\pm$ 1.3 \\
fit &  8.51&  -4.69& -27.60& -0.28\\
$\bar c c$ P1 & 30.9 $\pm$ 0.09 & -14.64 $\pm$ 0.07 & -110.6$\pm$ 0.07 & 0.07 $\pm$ 0.12 \\ 
fit & 30.96 & -14.6 & -110.38 & -0.19\\
\hline 
\label{tab_splittings}
\end{tabular} \end{center}
\caption{Splittings of the P-shell states from the ``unpolarized combination" (\ref{eqn_unpolarized}) , in ${\rm MeV}$. The values with error bars are from the Particle data Tables, while those in the next lines are from the fit to the first order expressions described in the text.}
\label{tab_split_compare}
\end{table}

\begin{figure}[h!]
    \includegraphics[width=8cm]{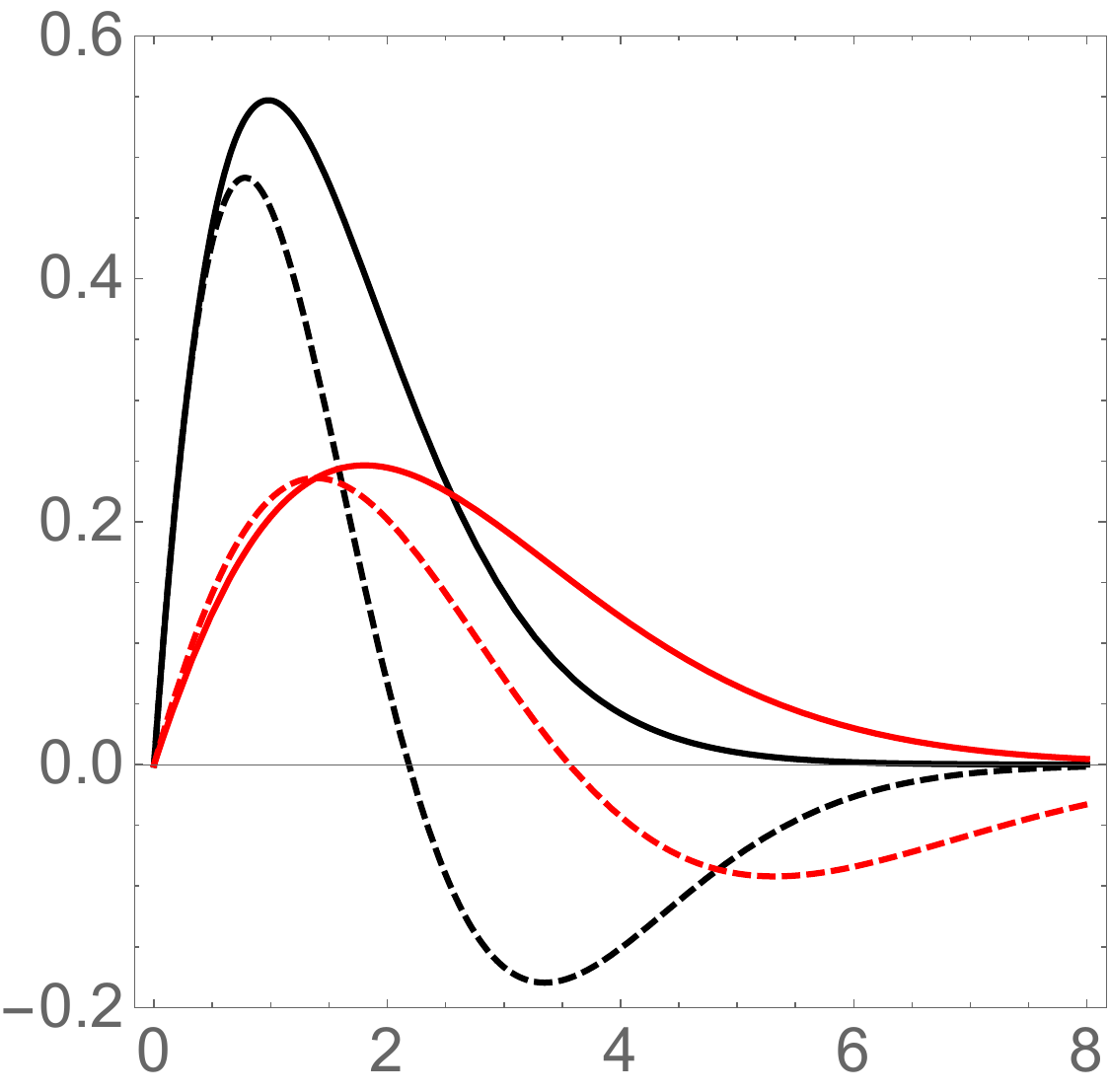} 
    \caption{The 1P wave functions of charmonium (black solid) and bottomonium (red solid curves)
    versus distance $r \, (GeV^{-1})$.
    The corresponding wave functions  in the 2P shell are shown by the dashed curves.}
    \label{fig_wfs_P.pdf}
\end{figure}

\subsection{Perturbative spin-dependent interactions}
\label{sec_pert}
The perturbative one-gluon exchange leads to the following well known
expressions for the spin-dependent potentials
\ba V_{SS}^{pert}&=& {4 \alpha_s \over 3} {2 \over 3 M^2} 4\pi \delta(r)\nonumber \\ 
 V_{T}^{pert}&=&{4 \alpha_s \over 3}{1 \over r^3 M^2} \nonumber\\
 V_{SL}^{pert}&=&{4 \alpha_s \over 3}{3 \over 2 M^2}{1 \over r^3}
\ea

In the earlier sections we discussed our procedure and parameters used
in the generation of the S-shell wave functions. The same procedure is followed
to generate the 1P and 2P wave functions, shown in
Fig.\ref{fig_wfs_P.pdf}. Note that, unlike the S-shell states, they vanish at the origin due to the added centrifugal potential. Therefore, the local perturbative spin-spin forces do not contribute to the masses of the $P$ shell states we now discuss.

The perturbative tensor and spin-orbit potentials are proportional to
the same integrals
\be 
\int dr r^2 |\psi_i(r)|^2 {1 \over r^3} 
\ee
Their contributions are different for the  three cases $i=1P, 2P \,\bar b b$ and $1P \,\, \bar c c$
under consideration. The tensor matrix elements (in MeV) are given by
\ba
\langle 1P \, \bar b b | V_{T}|  1P \, \bar b b \rangle =13.47 \nonumber \\
\langle 2P \, \bar b b | V_{T}|  2P\,  \bar b b \rangle =9.89 \nonumber \\
\langle 1P \, \bar c c | V_{T}|  1P \, \bar c c \rangle =19.32 \nonumber \\
\ea
They are to be compared to the phenomenological ones derived in the previous section, $13.2, 8.8, 40.4\, \rm MeV$, respectively. We note that
for both bottomonia quartets, the perturbative tensor force approximately 
reproduces the results of the phenomenological fit, but is
insufficient to explain the data for charmonium. The same can be said for
the spin-spin forces, which are basically zero within  errors. 

These results were  known since the 1970's, and subsequent
studies based on perturbative nonrelativistic QCD. In particular, in our paper 
in~\cite{Shuryak:2021fsu}, we have analysed a wider range of quark masses,
all the way from $b$ to light quarks, also documenting how the
magnitude of the spin-dependent forces evolve from perturbative to nonperturbative ones.

We will postpone the discussion of the perturbative spin-orbit forces,
to  later sections, when it will be discussed together with 
the nonperturbative effects.

\subsection{Spin-orbit forces induced by instantons and $\bar I I$ molecules} \label{sec_SL_inst}
The fundamental relations connecting the static potentials for quarkonium, wether
central or spin-dependent, are given in terms of Wilson lines. The spin-dependent terms come from the relativistic corrections 
$\sim (v/c)^2$ to the quark propagators,
with explicit  magnetic and electric gauge fields.  They were originally 
derived  in full generality in~\cite{FE}. For a review with  
applications in pnrQCD effective theory see~\cite{Brambilla:2004jw}. 
More explicitly, the spin dependent potentials are defined as
\begin{align} \label{eqn_pot_defs}
V_3  (\hat r^i \hat r^j) + {\delta^{ij} \over 3} V_4 &=
lim_{T\rightarrow \infty}\int_{-T/2}^{T/2}  dt \int_{-T/2}^{T/2}  dt'
\nonumber \\ 
&\times  \langle B^i(\vec x, t) B^j (\vec x' t') \rangle_W \nonumber \\
 {r^k \over r} 
{dV_1 \over dr} &= lim_{T\rightarrow \infty}  
{\epsilon_{ijk} \over 2} \nonumber \\ \times
\int_{-T/2}^{T/2}  dt & \int_{-T/2}^{T/2}  dt' ({t-t' \over T}) 
\langle B^i(\vec x, t) E^j (\vec x' t') \rangle_W \nonumber \\
{r^k \over r} 
{dV_2 \over dr} &= lim_{T\rightarrow \infty} {\epsilon_{ijk} \over 2} \nonumber \\ \times
 \int_{-T/2}^{T/2}  dt &\int_{-T/2}^{T/2}  dt' ({t' \over 2T})
 \langle B^i(\vec x, t) E^j (\vec x' t') \rangle_W
\end{align} 
with the  appropriate Wilson lines between the gauge fields assumed.
Here, the Wilson loop is rectangle $T \times r$, much like in the original
construction for the central potential. We will use these definitions
below, with  $\vec B,\vec E,W$ calculated for instantons or instanton molecules.

Note that  the spin-spin and tensor potentials are related to correlators of the
vacuum magnetic fields, where the quarks, via their magnetic moments,
interact with them directly. In the case of the spin-orbit interaction,
we need to evaluate the correlators of the
electric and magnetic fields. The appearance of the electric field
is related with the nonzero momentum of the quark. That is why
the spin-orbit potential is absent in S-shell hadrons, and present in  e.g. $L=1,P$-shell hadrons. Note also that
one unit of $L$ flips the $P$-parity of the states,
eliminating the possibility of their mixing with S- and D-shell states.

The general expression for the spin-dependent potentials is
\ba
V_{SD} &=&\big[  \vec S \vec L\big] \bigg( {1 \over 2 M^2 R} \bigg) \bigg( {d V_r \over dR} + 2 {dV_1 \over dr} +2 {dV_2 \over dr}\bigg) 
\nonumber \\
&+& \big[ 3 (\vec S_1 \hat \vec r)(\vec S_2 \hat \vec r)-(\vec S_1 \vec S_2)  \big] {1 \over 3 M^2} V_3 \nonumber \\
&+& [\vec S_1 \vec S_2]  {1 \over 6 M^2 } \vec\nabla^2 V_2 
\ea
(with a minor simplification related  to quarkonia made of quarks with {\em the same} mass $M$). Here, $V_r(r)$ is the central potential,

It is known since the 1980's, that the correlators of gluonic point-to-point operators
$\langle O(x) O(x')\rangle$ strongly depend on their quantum numbers.
The scalar $O_S=(G_{\mu\nu})^2$ and pseudoscalar  $O_{PS}=G_{\mu\nu} \tilde G_{\mu\nu}$ correlators show nonperturbative behavior up to very
large $Q^2\sim 20 \,GeV^2$, unlike e.g. the stress tensor operator $O_T$.
The reason for this is the dominance of the vacuum fields by the instanton configurations, (and thus strongly coupled to) $O_S,O_{PS}$  but having zero $O_T$, see details e.g. in \cite{Schafer:1994fd}. 

\begin{figure}
    \centering
    \includegraphics[width=7cm]{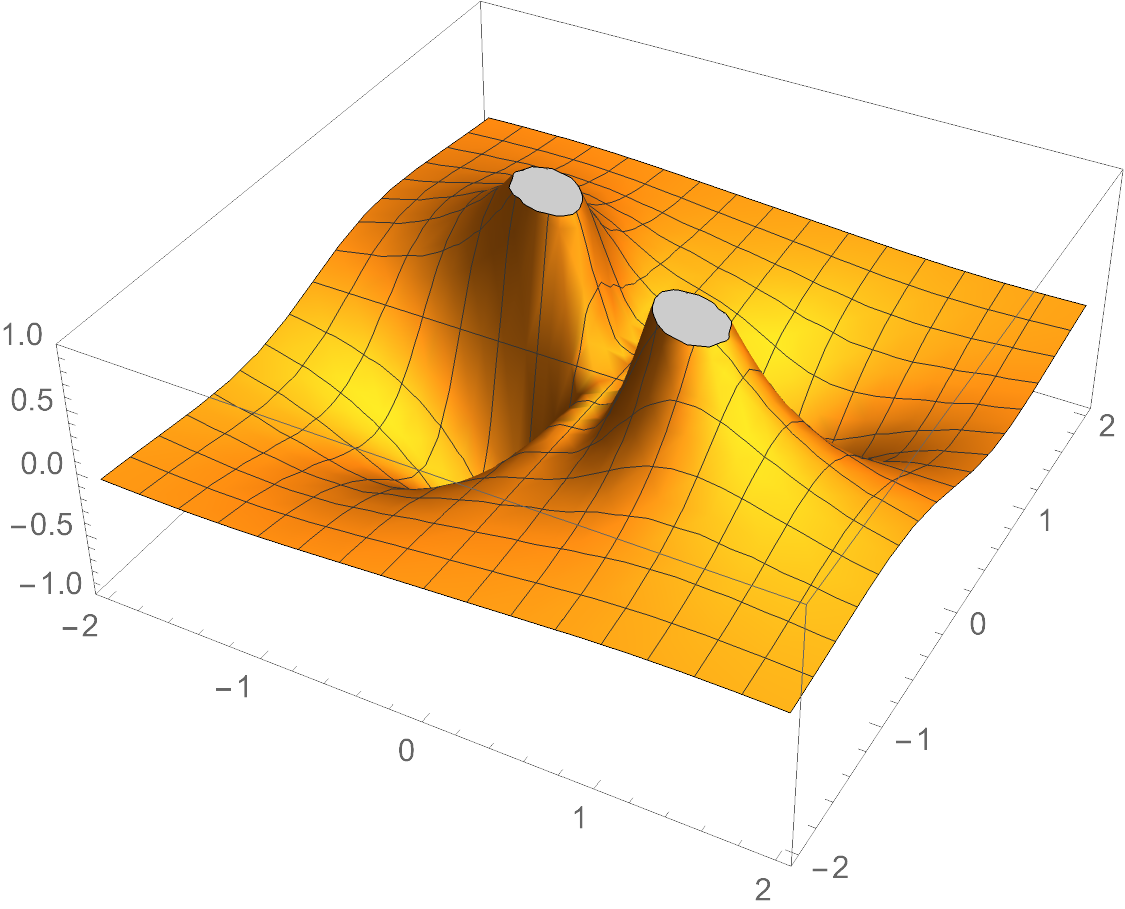}
    \caption{Distribution of the magnitude of the $P^1$ component of the  Pointing vector 
    in the  $x^1-x^4$ plane. }
    \label{fig_Pointing}
\end{figure}

These observations have direct relevance for the spin-orbit potentials
$V_1,V_2$ under consideration. 
Indeed, if $E^i$ and $B^j $ in their definitions
are at the same time moment and
 location, their vector product would be the
Pointing vector $$\vec P=[\vec E \times \vec B]$$
defining the  direction and magnitude of the energy flow  carried by the vacuum gauge fields. For any self-dual fields, 
 instantons included, all components of the stress tensor, 
 Pointing vector included, 
must $vanish$. 
This is of course  not the case 
for the ``molecular" configurations, made of a combination of self-dual and anti-selfdual parts. In particular, the Pointing vector $\vec P=[\vec E^a \times \vec B^a]\neq 0$  is nonzero.
Since the $\bar I I$ 
field configurations we discuss are still well localized, their energy can only flow from one place to another, in some closed lines,
which turns out to be approximately circular. An example of  the $\vec P$-distribution in some plane across the $\bar I I$ molecule, is shown in 
Fig.\ref{fig_Pointing}, with  $two$ visible and separate energy flow circles.

\begin{figure}[h!]
    \centering
    \includegraphics[width=7cm]{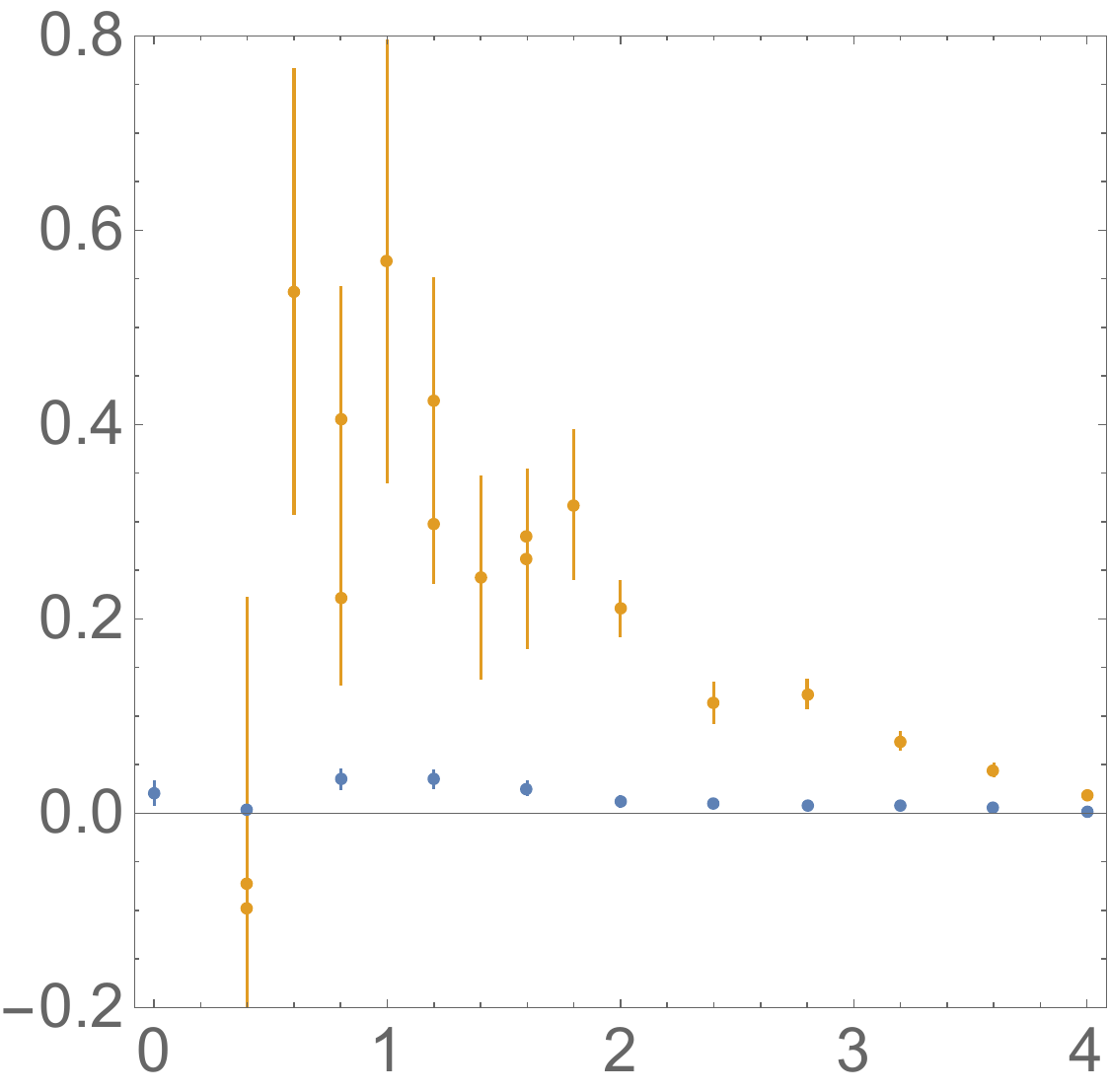}
    \includegraphics[width=7cm]{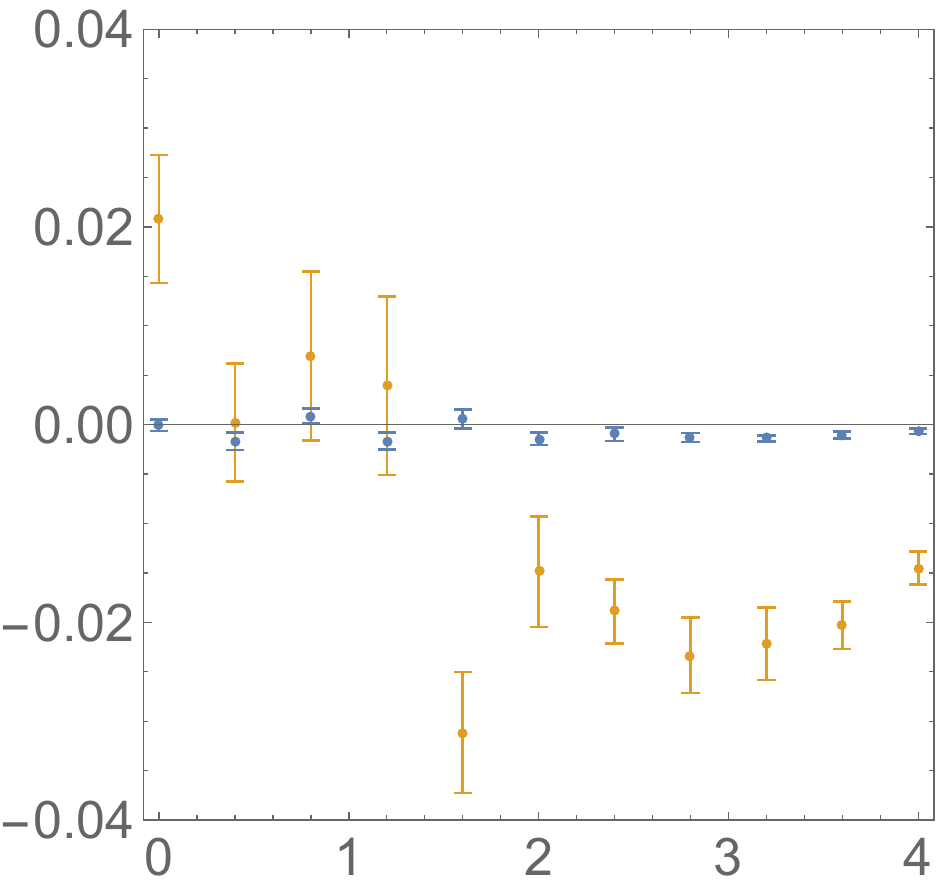}
    \caption{Derivatives of the potentials $dV_2/dr$ (upper plot) and $dV_2/dr$ (lower plot) in units of $n_{I}\rho^{2}$, as a function of the interquark distance $r/\rho$. They are calculated directly from the definitions 
    containing the
    correlators of the electric and magnetic fields (\ref{eqn_pot_defs}), as a function of $r/\rho$. The blue points are for an ensemble of single
    uncorrelated instantons, and the upper red points are  for an ensemble of  $\bar I I$ molecules.}
    \label{fig_BE_plot}
\end{figure}

The explicit evaluation of the correlators, that combine the
field strengths and Wilson lines, turns out to be technically 
complicated,  with a need to get significant statistics in
our Monte-Carlo approach to averaging over the space-time orientation 
and locations of molecules, see the discussion in Appendix \ref{sec_evaluation}. 
The results are plotted in Fig.\ref{fig_BE_plot}. Indeed, we note that the
  ``molecules" produce much larger  $EB$ correlators than the singe instantons. 
  We also note that $dV_2/dr$ in which the fields are on different quark lines,
  is significantly larger than $dV_1/dr$ (lower plot), in which $E$ and $B$
  act on the same quark. Finally, we note that  that the two potentials seem to
  have opposite signs.

The quantities plotted in Fig.\ref{fig_BE_plot} are for a single instanton or molecule, and given in units of the instanton size. To put them in absolute units, we need  to multiply  by a combination of the same dimension
 $n_{mol}\rho^2 $. As discussed in \cite{Athenodorou:2018jwu},
 the vacuum density of molecules should be obtained from the extrapolation to zero value of the gradient flow time,
 which at present is only known with significant uncertainties.
 $n_{mol}=8-12 \, fm^{-4} $. 
 If e.g. we use
$n_{mol}=9\, fm^{-4}, \rho=1/3\, fm$, this combination is $1 fm^2\approx (0.2\, GeV)^2$. When compared to the string tension $dV_r/dr=\sigma\approx (0.4\, GeV)^2$, the first factor in the expressions for the spin-orbit potentials  is $n_{mol}\rho^2/\sigma\approx 1/4$.

The evaluated correlators for the instantons and molecules are given in Fig.\ref{fig_BE_plot}. It follows that $dV_2/dr$ is small compared to $dV_1/dr$, which in turn is somewhat smaller than the string tension $\sigma$. We will return to the summary of the spin-orbit forces in section \ref{sec_SL_summary}.

\subsection{QCD strings contribution to spin-orbit potential}
\label{sec_SL_string}

Consider the standard Nambu-Goto string with end-points, carrying equal mass and spin,
\bea
\label{1}
S=&&\sigma_T\int_{-T/2}^{T/2}  d\tau\,d\sigma\, 
\sqrt{h}\nonumber\\
&&+\int_{-T/2}^{T/2}  d\tau\, \bigg(M\sqrt{{\dot{X}}^2}
-\frac 12 \sigma_{\mu\nu}
\frac{{\dot{X}}^\mu {\ddot{X}}^\nu}{{\dot{X}}^2}
\bigg)
\eea
with the volume element given by the determinant
\bea
h={\rm det}\bigg(\eta_{\mu\nu}\partial_aX^\mu\partial_bX^\nu\bigg)
\eea
The last boundary term is the spin factor which enforces Thomas precession along each of the end-point world lines~\cite{Strominger:1980xa,Polyakov:1988md}. Recall that 2 infinitesimal boosts are equivalent to 1 infinitesimal boost plus a Thomas precession. The latter is related to a Berry phase~\cite{Nowak:1990wu}.
Their contribution to the interaction potential is best seen by considering the classical solution to string EOM following from (\ref{1}),
\bea
\partial_\alpha\bigg(\sqrt{h} h^{\alpha\beta}\partial_\beta X^\mu\bigg)=0
\eea
subject to the spinless boundary conditions
at $\sigma=\pm 1$
\bea
\sigma_T\sqrt{h}\partial_\sigma X^\mu\pm M\partial_\tau\bigg(\frac{\dot{X}^\mu}{\sqrt{\dot{X}}^2}\bigg)=0
\eea
The rotating string in the 12-plane is such a solution~\cite{Sonnenschein:2014bia} (see also references therein), with the embedding
\bea
\label{2}
X^{\mu}=(\tau, 
r\sigma\, {\rm cos}(\omega\tau),
r\sigma\, {\rm sin}(\omega\tau), 0)
\eea
which solves the EOM, and for which the boundary condition reduces to
\bea
\label{SIGMAT}
\frac{\sigma_T}\gamma =\gamma M\frac{v^2}r
\eea
The end-point velocity is $v=\omega r$ and 
the Lorentz contraction factor is fixed by $1/\gamma^2=1-v^2$. In (\ref{SIGMAT}) the tension in the string balances the centrifugation, with the correct relativistic gamma factors.

From the Noether currents following from Poincare symmetry, the energy is
\bea
\label{P0}
P^0=2\gamma M+\frac{\sigma_T 2r}{v}\int_0^v\frac{dx}{\sqrt{1-x^2}}-\frac 12
(\sigma_1^z+\sigma_2^z)\gamma^2v^2\omega\nonumber\\
\eea
and angular  momentum
\bea
\label{JZ}
J^z=2\gamma Mr^2+\frac {2\sigma_T \omega}{v^3}
\int_0^v\frac {x^2\,dx}{\sqrt{1-x^2}}
+\frac 12 (\sigma_1^z+\sigma_2^z)\nonumber\\
\eea
 The spin contribution to the rotating string can be recast using the boundary constraint (\ref{SIGMAT})
 \bea
 \label{VTSL}
 V_{TSL}(r)=-\frac 12
(\sigma_1^z+\sigma_2^z)\gamma^2v^2\omega
=-\frac{1}2\frac {S^zL^z}{M^2 r} \frac{dV_C(2r)}{d2r}\nonumber\\
\eea
with $L^z=mvr$ and $S^z=\frac 12(\sigma_1^z+\sigma_2^z)$ and the
confining part of the central potential 
$V_C(r)=\sigma_T r$. 
It is the $scalar$ spin-orbit interaction 
arising from Thomas precession, the resultant of
successive  boosts along the locus of the massive end-points with attached spins. This contribution from the confining string, was originally noted in~\cite{Buchmuller:1981fr,Gromes:1984ma,Pisarski:1987jb}.

One should further note that the contribution from the NG string receives quantum corrections, resummed in the Arvis potential~\cite{Arvis:1983fp}
\bea
\label{ARVIS}
V_C(r)=&&\bigg(\sigma_T^2r^2-\frac{D_\perp}{24\alpha'}\bigg)^{\frac 12}=\sigma_T(r^2-r_c^2)^{\frac 12}\nonumber\\
\approx&&\sigma_Tr-\frac{\pi D_\perp}{24 r}-\frac {\pi^2}{2\sigma_Tr^3}\bigg(\frac{D_\perp}{24}\bigg)^{\frac 12}
\eea
with the number of transverse dimensions $D_\perp=2$, $\alpha'=l_S^2=1/2\pi\sigma_T$ and the critical string length is
$$r_c=\pi\bigg(\frac{\alpha'D_\perp}{6}\bigg)^{\frac 12}\rightarrow \frac{\pi l_S}{\sqrt 3}$$
The large distance expansion yields the familiar linear confining part plus the universal ``Luscher term".
The Arvis potential shows that the string description no longer applies for small distances $r<r_c$, 
in contradiction to the phenomenological Cornell-type potentials.


\subsection{Quantifying all  spin-orbit forces} \label{sec_SL_summary}
After the derivation of the spin-orbit potentials induced by the vacuum solitons and string-based contributions in the preceding section, 
we  now proceed to evaluate numerically  these contributions,
and compare the results with fits to data from the previous sections. We recall that those are deduced from
the observed mass splittings 
of three "quartets" of quarkonia,  $P1$ and $P2$ in $\bar b b$ and  $P1$ in  $\bar c c$ shown in the upper raw of Table \ref{tab_LS_string}.

The theoretical evaluations are done by  averaging $V_{LS}$ with appropriate wave functions. It
starts with the perturbative potential approximated in a form $V_{LS}^{pert}=0.64/r^3$ (the second row). Note that the first
two values (for bottomonium) significantly overshoot the
data fit, while for charmonium they are lower.
The perturbative estimates were done  many times before, with the
suggestion to incorporate smaller $\alpha_s$ coupling for bottomonia, and 
reproduce the data. Our comment here is that the coupling
should not depend on the quark flavor, but depends on the distance $r$, and in fact $P2$ in $\bar b b$ and  $P1$ in  $\bar c c$ wave functions have the same sizes.

The effects of the vacuum solitons is estimated for $\bar I I$
molecules, which are much larger than that for single instantons, see Fig.\ref{fig_BE_plot}. Also we  neglected the smaller 
contribution from $dV_1/dr$. The averaging 
of ${1\over r}{dV_2 \over dr}$ is carried  with the wavefunctions, see the results in  the third row of the summary Table. For both $P1$ shells, the contributions are about half the phenomenological values. (The smaller  value for the $P2$ bottomonium state is related with the nodes
of the wave function where the potential is the largest.)

If we add the perturbative and nonperturbative contributions,
the theoretical answer would overshoot the experimental one, by
about a factor of 1.5-2. This is the version of  the ``spin-orbit puzzle", in this  analysis.

As we suggested earlier and discussed in the preceding section, perhaps the problem can be cured by the long-distance
Thomas term which is negative in the string formulation. 
Let us quantify this suggestion.
We will do so using two limiting forms of the potentials
\ba
V_{SL}^{string}&=&{\sigma \over M^2 r}, \nonumber \\
V_{Arvis}&=&\sigma r \sqrt{1-{\pi \over 6 \sigma r^2} }, \nonumber \\
V_{SL}^{Arvis}&=&{1 \over 2 M^2 r}{dV_{Arvis} \over dr }
\ea
(Recall that the Arvis potential and the corresponding $LS$ potential, taking care of the string vibrations, should only be used
for distances where the argument of the square root is positive.)

Since the value of the string tension at large distances is 
uncertain, we will not use the fitted value in the Cornell potential,
but just put the traditional number $\sigma=(0.4 \, GeV)^2$. The result of averaging with the wave functions of the $P1$ shell are in the bottom two rows of the Table \ref{tab_LS_string}. 

Within the  existing uncertainties, the string tension term is thus comparable to the ``molecular"
contribution we found. Subtracting it from the latter, we find a phenomenologically acceptable result for $\bar c c$. Yet for both $\bar b b$ cases, the sum of the theoretical contribitions is still larger than the phenomenological values. As we already mentioned,
this  can possibly be cured by the reduction of $\alpha_s$ used in the perturbative part.

\begin{table}[h!]
    \centering
    \begin{tabular}{cccc}
\hline
  & $1P \, \bar b b$ &  $2P \, \bar b b$ &  $1P \, \bar c c$ \\  \hline
   fit to data & 13.6 & 9.3 & 35. \\
 $V_{LS}^{pert}$ & 20.2 & 14.8 & 29.0\\
  $V_{LS}^{molec}$ & 7.26 & 3.75 &  16.5 \\
  $V_{LS}^{string} $ & {\bf -4.77} & {\bf -3.40} & {\bf-12.9 }\\ 
  $V_{LS}^{ Arvis }$   & {\it -1.68} & {\it -0.87} & {\it -11.2} \\ 
 \hline         
    \end{tabular}
    \caption{ Matrix elements (in MeV) of the spin-orbit interactions, for the three quarkonia ``quartets" indicated.
    The upper row is our fit to the quarkonia data. The next is the perturbative contribution, and that due to   the ``instanton molecules" in the QCD vacuum. The bottom
    two rows (bold) are the contributions of the string-related potentials with Thomas precession, from a linear term and Arvis generalization, respectively.  }
    \label{tab_LS_string}
\end{table}

\section{Summary}
\label{sec_CON}
This paper follows the conjecture made in our paper~\cite{Shuryak:2021fsu}, that the nonperturbative central potentials at intermediate distances 
$r\in [0, 2/3\, fm]$, can be described by ensemble of instantons
in the QCD vacuum. Indeed, precisely  in this region, the  description
via QCD strings is probematic, as Arvis' resummation of the string
quantum oscillations shows. 

Here we introduced a certain ansatz for the description of strongly correlated instanton-antiinstanton pairs, or ``molecules". 
The static potential following from a temporal Wilson loop going through a
``molecular" $A_\mu^a$, yields the same linear function of the distance
as for instantons, in the most important intermediate distance range, see Fig.\ref{fig_WW_plot}.

The main objectives of the current paper are the phenomenological analysis and the theoretical evaluation of the {\em spin-dependent potentials}. We started our analysis of the
spin-spin potential $V_{SS}$, which is at the origin of the  splittings in the S-shell quarkonia, such as $\Upsilon-\eta_b$.  In particular, we addressed two of these in the  $\bar b b$ family, and the lowest in $\bar c c$.

It is well known, that the perturbative $V_{SS}$ is local, and therefore
its matrix elements are proportional to the squared wave functions at the origin. The non-perturbative $V_{SS}$, is related to the correlator of magnetic fields, and is also very much localized at small distances, but it is $not$ local. Unlike the central potential, $V_{SS}$ due to ``molecules" turns out to be much larger than for single instantons. 

When comparing the corresponding
matrix element to data,  we have concluded that about $\frac 13$
of the spin splitting is nonperturbative. Note that this holds  even in bottomonia, hadrons of the smallest size available. 

Proceeding to quarkonia in the $P$-shell, we looked at the $P1,P2$ quartets of states $\chi_0,\chi_1,\chi_2,h$ in bottomonia and  
$P1$ in charmonia. Those three quartets are far enough from  thresholds
of meson-meson states, and can be considered to be ``clear" quarkonia. We made explicit 
their spin/orbit wave functions, and derived their angular/spin averages for all three
potentials, $V_{SS},V_T,V_{SL}$. Minimizing the chi-squared of the averages, we obtained
a fit to their matrix elements, leading
to an excellent description of all the splittings in the three quartets, see Table \ref{tab_splittings}. It turns out that two are comparable, but the spin-spin
effect is so small that it is basically {\em not seen} inside the error bars.

 This is hardly surprising since the $P-shell$ wave function vanishes at the origin due to the centrifugal potential. A well know fact that perturbative local term vanishes. The non-perturbative effect (which we calculated as being due to
 ``molecular" magnetic fields) also turned out to be rather short range, giving also a small effect. 

In contrast to that, the spin-orbit potentials induced by ``molecules" due to the electric-magnetic $[\vec E \times \vec B]$ correlators are neither small (as for single instantons) nor near-local. They produce effects comparable to large distance Thomas precession effect, and even the perturbative 
spin-orbit potential. As the matrix elements  listed in Table \ref{tab_LS_string} shows, the
 contributions of the ``instanton molecules" and the string-based Thomas precession effect, 
 nearly cancel each other.
In summary, we have not found any ``spin-orbit puzzle" in the
$P$-shell quarkonia.

A theoretical question on whether instanton-induced and string-induced effects can be smoothly
connected still remains. Phenomenologically, one can approach this issue by
focusing on larger-size mesons (lighter quarks or higher principle quantum numbers). 
Theoretically, one may try to push theoretical formulae to higher accuracy, from both ends.
 
The spin-orbit puzzle however still does exist for
 $P$-wave (light quark) baryons.
 Decades ago,  Isgur and Karl \cite{Isgur:1978xj}
obtained a
remarkable phenomenological success of their description of the $P$-shell   odd parity baryons,  using  only spin-spin and tensor interactions induced by the one-gluon exchange, while the 
  spin-orbit interaction was  altogether omitted. Their proposed solution to 
 the spin-orbit puzzle was a $conjecture$ that negative Thomas precession 
contribution at large distances can be large enough to cancel all other terms, including
the perturbative one. 
 
 Of course, what happens with the spin-dependent forces in the $P$-shell baryons still needs to be investigated. In our recent work 
 \cite{Miesch:2023hvl} we made some steps to this goal, 
 rederiving the explicit spin-isospin wave functions
 in spinor-tensor form, and obtained the
 phenomenological matrix elements for $V_{SS},V_{T},V_{SL}$
 from current data on five nucleon splittings. In agreement with 
 Isgur and Karl, we indeed observed that the $V_{SL}$ matrix element
 seems to be null, within errors
 (unlike in heavy quarkonia studied in the current work.). So, the spin-orbit puzzle 
 is indeed real for light baryons.

In view of this, let us end the paper with some suggestions for future works. 
 One issue is that for hadrons made of light quarks 
 we do not have  wave functions because one does not have a reliable nonrelativistic Schroedinger equation. This in principle can be 
 remedied by going to the light front formulation we worked out
 elsewhere, in which heavy and light quarks are treated by similar Hamiltonians and
 wave functions.

One can also investigate what happens with heavy-quark baryons. We also evaluated in this work the three-quark interactions in baryons
 using three Wilson lines, in a way similar to the central 
 quark-antiquark potentials.
Unfortunately, we do not have the experimental detection of  heavy-quark baryons $ccc$
and $bbb$ yet. Yet one can proceed in theory, as we did in 
 \cite{Miesch:2023hvl}, using the hyperdistance approximation.
There are multiple efforts to study heavy quark baryons on the lattice. We are not yet aware of studies of the spin-dependent forces or the $P$-shell states, but it can be done.

Let us end the paper by emphasizing once more, its main goal: it is important
to connect  hadronic spectroscopy with available  models of the vacuum structure. A specific
point which stands out in this work is that the spin-spin and spin-orbit
forces we see in hadronic splittings do tell us about
magnetic-magnetic and electric-magnetic field correlators
in the vacuum. They are nontrivial and, as we have demonstrated, they may be approximated using correlated pairs of  topological solitons.



\vskip 1cm

\noindent{\bf Acknowledgments.\,\,}This work is supported by the Office of Science, U.S. Department of Energy under Contract No. DE-FG-88ER40388, and,
in part under contract no. DE-SC0023646,  within the framework of the Quark-Gluon Tomography (QGT) Topical Collaboration.

\appendix
\section{4-dimensional rotations} \label{sec_4d_rotations}

One can either rotate the $I\bar I$ molecule 
fields to arbitary orientations, as is the case in
the vacuum, and keep the Wilson lines oriented along the (Euclidean) time, or 
shift and rotate instead the Wilson lines. We have used the former approach,
rotating the vector $A_\mu$ itself as well as its argument. Therefore, by electric and magnetic
fields we use those in the quark rest frame.

The $SO(4)$ rotation group is known to have 6 angles,  best described as a convolution of two $SO(3)$ rotations.  We use the so called Van Elfrinkhof formula which defines an $SO(4)$ rotation as the product
of two matrices 
\ba
\begin{bmatrix} 
	a & -b & -c & -d \\
	b & a & -d & c\\
	c & d & a & -b \\
 	d & -c & b & a \\
	\end{bmatrix}.
 \begin{bmatrix} 
	p & -q & -r  &-s \\
	q & p & s & -r\\
	r  & -s & p & q \\
     s & r & -q  & p\\
	\end{bmatrix}
\ea
built from random unit 4-vectors $(a,b,s,d)$ and $(p,q,r,s)$. The formula was historically based on quaternion expressions due to Cayley (1854).
It rotates the original 4 coordinate unit vectors to 4 others, all unit vectors rotated to unit ones which are mutually orthogonal. 
Obviously they belong to the $SO(4)$ group, so that $\hat R^T=\hat R^{-1}$.

\section{Evaluation of correlators} \label{sec_evaluation}
We now  outline the procedure we followed. The general setting has already been explain in Fig.\ref{fig_cube}. Now, suppose the 4d density of the objects we include is $n_{mol}=1/V_4$.
This defines an ``elementary cell" as a cube, of size $R_{cell}=(V_4)^{1/4}$, in which on average
one solitonic object is located. We have followed the ``single cell" approximation, including one
solitonic object  in the cell only,  pierced by Wilson lines. 

The other solitonic objects (e.g. two more molecules shown in Fig.\ref{fig_cube}) are {\em not included}. This 
should be possible, provided their contribution to the gauge fields decay   sufficiently fast with the distance, which is  indeed  the case. (Recall that e.g. $\vec B^2$ decreases from 
the instanton center as $1/r^8$.)

Let us define two
frames, in which the coordinates are referred to as 4-vectors $y^\mu$
and $x^\mu$. The former is the one in which the gauge field configurations have the standard form, e.g. the ansatz (\ref{eqn_ansatz}) for $\bar I I $ molecules. The latter is 
the frame in which the quarks are at rest, but the fields are rotated.
The relation between them is defined by $y=\hat R.x$ where $\hat R$ is the random
rotation matrix belonging to the $SO(4)$ orthogonal group.
As the correlator is in Euclidean space-time,  such rotations are
substitute for Lorentz transformations.
Note also that it is in standard notations,
as the ``inverse" rotation matrix to the
 rotation $y\rightarrow x$,  by which the lower-index fields $A_\mu,G_{\mu\nu}$ are rotated. The coordinates
$x^\mu$ are rotated by the "direct" matrix $\hat R^{-1}$,
leaving scalars like $A_\mu dx^\mu $ unchanged.
Finally, note that the fields $\vec B,\vec E$ in the correlators are
those defined in the frame of the quarks, $after$ the evaluation of the rotated $G_{\mu\nu}$
of the molecule.

We have used the Monte-Carlo approach to  average over (i) the location of the molecule in the box); (ii) the orientation of the randomly rotated molecule, in total  3+6=9 parameters. 
parameters). So we either relied on Mathematica integration in 9d, or on 
a direct Monte-Carlo approach where we assigned random values to all variables.
The second method is straightforward but 
is very far from efficient, and the error bars shown follow from fluctuation of the evaluation. 
 We were only able to reach such statistical significance, using parallelization tools in Mathematica and a MacBook 
running  with 32 cores.

Let us now provide some possible caveats to our approach which may be raised. One may question the accuracy of a ``single cell" approximation
at large distances , $r>R_{cell}$, since objects outside the primary box may then be closer to the Wilson lines than the one in the cell. However, We do not think that this issue  can seriously affect the results.

Let us mention
the specific numerical values used. If $n_{mol}=9\, fm^{-4} $ the cell size
is $R_{cell}=fm/\sqrt{3}=\sqrt{3}\rho$. The potentials have been evaluated till distances between quarks $r\sim 4/3\, fm = 4\rho$. Yet these potentials are generally decreasing 
with distance. Furthermore, they are
only used in matrix elements, 
with extra $1/r$ and with heavy quarkonia wave functions squared, dominated by distances of order $r\sim2\rho$.
And, last but not least, as argued in the text, at large $r$ the Thomas term is expected
to provide a negative $-\sigma$ contribution, overshadowing the calculated $dV_1/dr,dV_2/dr$ anyway.
More generally, the instanton-based and string-based approaches are supposed to be
best used at small and large distances, respectively. How to connect them at intermediate $r$ remains an open problem.

\section{Quarkonium P-shells: phenomenological inputs}
\label{sec_Pshell}
In this work we  focus on {\em three sets} of $L=1$ or $P-shell$ quarkonium states,
namely $P1,P2$ sets in $\bar b b$ and $P1$ in $\bar c c$. Only these three sets have masses which are
known well enough, and also they are not too close to the mesonic decay thresholds.
With two spins and 3 components of orbital momentum, each consists of $2^2\cdot 2^2\cdot 3=12$ states. Physical states with particular $J^{PC}$ quantum numbers
are referred to as $\chi_2,\chi_1,\chi_0,h$, where the  first three have $J$ given as
a subscript, and $h$ has $S=0,J=1$. Their masses are all given in the Particle Data Tables (2022 version) we used, with current error bars. 

Let us start with the ``observed splitting" of these states from that of the ``unpolarized combination"
\be \label{eqn_unpolarized}
M_{unpolarized}={5M_{\chi_2}+3M_{\chi_2}+M_{\chi_0}+3M_h \over 12 }
\ee
in which all (first order) spin-dependent contributions should vanish, see Table \ref{tab_splittings}. Note that in order to compare the splitting in charmonium,
the third line, to bottomonium ones (lines 1 and 2) one has to multiply these
data by $(m_c/m_b)^2 \sim 1/10$. After it is done, one finds that the spin-dependent
forces monotonously decrease with sizes of the state, from the most compact
$\bar b b$  P1 down the table. 

\begin{table}[h!] \begin{center} \begin{tabular}{c c c c c }
\hline
&                 $\chi_2$ & $\chi_1$ &  $\chi_0$ & h   \\
\hline
$\bar b b$  P1 & 12.5 $\pm$ 0.5 & -6.9 $\pm$ 0.5 & -40.3 $\pm$ 0.6 & -0.4 $\pm$ 0.8 \\
$\bar b b$  P2 &  8.5 $\pm$ 0.7  & -4.7 $\pm$ 0.7 & -27.6 $\pm$ 0.8 & -0.3$\pm$ 1.3 \\
$\bar c c$ P1 & 30.9 $\pm$ 0.09 & -14.64 $\pm$ 0.07 & -110.6$\pm$ 0.07 & 0.07 $\pm$ 0.12 \\ \hline 
\label{tab_splittings}
\end{tabular} \end{center}
\caption{Splitting of P-shell states from ``unpolarized combination" (\ref{eqn_unpolarized}) , in $\rm MeV$}
\label{tab_splittings}
\end{table}

The wave functions of the states are generally known, but for completeness let us also
present here their explicit spin-orbit parts using the spin-tensor notations
in Mathematica. (We already used it in a much more complicated case of $P-shell$
nucleons in~\cite{Miesch:2023hvl}, where the spin-isospin-tensors have 6 indices). Two quark spin components form a natural basis for
the spin-tensor with indices $s1,s2=1,2$. In this case there are only two spin structures, $$|\uparrow \uparrow \rangle,(|\uparrow \downarrow\rangle +\downarrow \uparrow\rangle)/\sqrt{2}, |\downarrow \downarrow\rangle) $$ for spin $S=1$ and $$(|\uparrow \downarrow\rangle -\downarrow \uparrow\rangle)/\sqrt{2}$$ for spin zero, $S=0$, in $h$
component of the quartet. In tensor notations $\uparrow=\{0,1 \}, \downarrow=\{1,0 \},$.

The Clebsch-Gordon coefficients and angular
functions are given explicitly by
\ba
\chi_{2,2}&=& \{ \{   -{e^{i\phi} sin(\theta)  \over 2} \sqrt{3 \over 2\pi }, 0 \} , \{0,0\}\}\nonumber \\
\chi_{2,1} &=& \{ \{ cos(\theta)\sqrt{3 \over 2\pi }/2,
-{1 \over 4}\sqrt{3 \over 2\pi }e^{i\phi} sin(\theta)\}, \nonumber \\
\{ &-&{1 \over 4}\sqrt{3 \over 2\pi }e^{i\phi} sin(\theta)\},0\}\}
\nonumber \\
\chi_{2,0} &=& \{\{ {e^{-i\phi} sin(\theta)\over 4 \sqrt{\pi}}, {cos(\theta) \over 2 \sqrt{\pi}} \}, \nonumber \\
\{ &+& {cos(\theta) \over 2 \sqrt{\pi}}, {-e^{i\phi} sin(\theta)\over 4 \sqrt{\pi} } \}\}
\nonumber \\
\chi_{1,1} &=& \{ \{ cos(\theta) \sqrt{3 \over 8\pi} ,{ e^{i\phi} sin(\theta)\over 4} \sqrt{ 3 \over 2\pi} \},\nonumber \\
\{ &+&  {e^{i\phi} sin(\theta)\over 4} \sqrt{ 3 \over 2\pi}, 0\}\} \nonumber \\ \ea
\ba
\chi_{1,0} &=& \{ \{ {-e^{i\phi} sin(\theta)\over 4}\sqrt{3 \over \pi} , 0 \},\nonumber \\
 &+& \{ 0, {e^{i\phi} sin(\theta)\over 4}\sqrt{3 \over \pi} \}\}\nonumber \\
\chi_{0,0} &=& \{ \{ {-e^{-i\phi} sin(\theta)\over 2\sqrt{2\pi}},-{cos(\theta) \over  2\sqrt{2\pi}} \}, \nonumber \\
 &-& \{ {cos(\theta) \over  2\sqrt{2\pi}}, {e^{i\phi} sin(\theta)\over 2\sqrt{2\pi}} \}\}\nonumber \\
h_0 &=& \{ \{0,{\sqrt{3 \over 8\pi} cos(\theta}  \},\nonumber \\
&-& \{ \{\sqrt{3 \over 8\pi} cos(\theta),0 \}\}   
\ea
All the wave functions times their conjugates, integrated over $sin(\theta) d\theta d\phi $ give 1. The  mutual products of different pairs give zero. 

The spin operators can be either used with the corresponding indices, or transformed into full dimension (here $4\times 4$) matrices. An example of the former for $\vec S_1 \vec S_2$ with all indices open is defined by the Mathematica statement 
\ba && Table[Sum[\sigma[k][[i1,j1 ]]*\sigma[k][[i2,j3 ]]/4, \{k,1,3\}],
\nonumber \\
&& \{i1,1,2\},\{i2,1,2\},\{j1,1,2\},\{j1,1,2\} ] \ea 
where $\sigma[k]=PauliMatrix[k]$. The resulting 4-index object need to be convoluted with the wave functions, two-index each, etc.

\bibliography{main}
\end{document}